\documentclass[prb,twocolumn,superscriptaddress,floatfix,showpacs]{revtex4}
\usepackage{amsmath,epsf,dcolumn}

\newcommand{\BIG}[2]{\mbox{$\left#1\vbox to #2{}\right. $}}
\newcommand{\be}{\begin{equation}}
\newcommand{\ee}{\end{equation}}
\newcommand{\bea}{\begin{eqnarray}}
\newcommand{\eea}{\end{eqnarray}}
\begin{document}
      
\title{Constraint-based, Single-point Approximate Kinetic Energy  %
Functionals }

\author{V.V.~Karasiev}
\email{vkarasev@qtp.ufl.edu}
\affiliation{Centro de Qu\'{\i}mica, Instituto Venezolano de Investigaciones
Cient\'{\i}ficas, Apartado 21827, Caracas 1020-A, Venezuela}
\affiliation{Quantum Theory Project, Departments of Physics and 
of Chemistry, University of Florida, Gainesville, FL 32611}
\author{R. S.~Jones}
\affiliation{Department of Physics, Loyola College in Maryland,
4501 N.\ Charles Street, Baltimore, MD  21210}
\author{S.B.~Trickey}
\email{trickey@qtp.ufl.edu}
\affiliation{Quantum Theory Project, Departments of Physics and
of Chemistry, University of Florida, Gainesville, FL 32611}
\author{Frank E.~Harris}
\affiliation{Department of Physics, University of Utah, Salt Lake City, UT}
\affiliation{Quantum Theory Project, Departments of Physics and
of Chemistry, University of Florida, Gainesville, FL 32611}

\date{03 September 2008, Version E2b}

\begin{abstract}

We present a substantial extension of our constraint-based approach
for development of orbital-free (OF) kinetic-energy (KE) density
functionals intended for the calculation of quantum-mechanical forces
in multi-scale molecular dynamics simulations.  Suitability for 
realistic system simulations requires that the OF-KE functional 
yield accurate forces on the nuclei yet be relatively
simple.  We therefore require that the functionals be based on
DFT constraints, 
local, dependent upon a small number of parameters fitted to a
training set of limited size, and applicable beyond the scope of the
training set.  
Our previous ``modified conjoint'' generalized-gradient-type
functionals were constrained to producing a
positive-definite Pauli potential.  Though distinctly better than several
published GGA-type functionals in that they gave semi-quantitative
agreement with Born-Oppenheimer forces from full Kohn-Sham results,
those modified conjoint functionals suffer from unphysical
singularities at the nuclei.  Here we show how to remove such
singularities by introducing higher-order density derivatives.  We
give a simple illustration of such a 
functional used for the dissociation energy as a function of bond length
for selected molecules.


\end{abstract}
                        
\maketitle

\section{Introduction}

Simulation of the structure and properties of complicated materials is
a demanding task, particularly away from equilibrium, for example, in
the simultaneous presence of solvents and mechanical
stress.  Though the present results are not limited to it, our
motivating problem has been tensile fracture of silica in the presence
of water.

For such problems, quantum mechanical treatment of the reactive zone
is essential at least at the level of realistic
Born-Oppenheimer (B-O) forces to drive an otherwise classical molecular
dynamics (MD) or molecular mechanics (MM) calculation.  Computational
cost then leads to a nested-region strategy.  Internuclear forces in
the reactive zone are obtained from an explicitly quantum mechanical
treatment. Forces between other nuclei are calculated from classical
potentials. Such partitioning is called multi-scale simulation in the
computational materials community and QM/MD (or QM/MM) methodology in
computational molecular biology.

The QM calculation is the computationally rate-limiting step.  QM
approximations good enough yet computationally fast on the scale of
the MD algorithms are therefore critical.  Both the inherent form and
the growing dominance of density functional theory (DFT) for
describing molecular, bio-molecular, and materials systems make it a
reasonable candidate QM.  Despite advances in pseudopotentials and
order-$N$ approximations, however, solution of the DFT Kohn-Sham (KS)
problem is too slow computationally to be fully satisfactory. An
alternative, Car-Parrinello dynamics \cite{CarParrinello},
 does not guarantee that the
motion is restricted to the B-O energy surface. 

Thus there is continuing need for methods which yield essentially full
DFT accuracy at significantly lower computational cost.  In response,
we and co-workers proposed and demonstrated a Graded Sequence of
Approximations \cite{GSA} scheme.  Its essence is use of a simple,
classical potential for the majority of MD steps, with periodic
correction by calibration to forces obtained from more accurate but
slower methods.  An example Graded Sequence of Approximations would be:
(1) classical potential; (2) simple reactive (charge re-distribution)
potential \cite{SM,Goddard,Goddard2}; (3) Orbital-free (OF) DFT, the
subject of this paper; (4) Quasi-spin density DFT \cite{QSDA} (a way
to approximate spin-dependent effects at the cost of
non-spin-polarized KS-DFT); (5) Full spin-polarized DFT (the level of
refinement ultimately required for bond-breaking).

In this hierarchy, a large gap in computational cost separates reactive
potentials and quasi-spin density DFT.  Since the cost of conventional
KS calculations comes from solving for the KS orbitals, an obvious
candidate to fill the gap is OF-DFT.  The long-standing
problem is a suitable OF approximation to the kinetic energy (KE).  
Background about the problem
and a detailed description of our first OF-KE functionals were
reported in a paper addressed to the computational materials science
community \cite{Perspectives}, with a more didactic survey in Ref.\
\onlinecite{Signpost}.  The present analysis focuses 
on identifying the causes of limitations of those functionals and
ways to eliminate those limitations.

\section{Background Summary}

Construction of an accurate, explicit total electronic kinetic energy
density functional
$T[n]=\langle\Psi|\hat{T}|\Psi\rangle$ for a many-electron system in
state $|\Psi\rangle $ with electron number density $n$ is an unresolved task 
\cite{Hohenberg-Kohn,Yang-Parr-book,DreizlerGrossBook}. The
Coulomb virial theorem suggests that the task is equivalent
to seeking the total energy functional itself.  The 
Kohn-Sham KE is thus a more attractive target for multiple reasons. 
Of course,
the appeal of OF-DFT predates 
modern DFT, as witness the 
Thomas-Fermi-Dirac \cite{Thomas,Fermi} and von Weizs\"acker 
\cite{Weizsacker} models.   There has
been considerable activity more recently. A review with extensive
references is given in Ref.~\onlinecite{LudenaKarasiev2002}.  Other
relevant work is that of Carter and co-workers; a helpful review
with many references is Ref.~\onlinecite{WangCarter2000}.  More
recent developments include, for example, 
Refs.~\onlinecite{Zhou06,Garcia0807,Perdew07,GarciaCervera08, %
Ghiringhelli08,Eek06} as well as our own
work already cited.

Distinct from most other recent efforts, our approach is 
to construct \emph{one-point, i.e., local} approximate KE
functionals specifically for MD computations.
We insist on constraint-based forms and parameters,
that is, satisfaction of known exact results for
positivity, scaling, and the like.  We are willing, as
needed, to simplify the search by requiring only that the functional 
give adequate interatomic forces, not total
energies (and certainly not KS band structures nor general linear response).

To summarize basics and set notation, we first note that
except as indicated otherwise we use Hartree atomic units.
The Kohn-Sham \cite{KS}
kinetic energy $T_{\rm s}$, the major contribution to $T$, is defined
 in terms of the KS orbitals:
\begin{eqnarray}
T_{\rm s}[\{\phi_i\}_{i=1}^N]
&=&\sum_{i=1}^{N}\int 
\phi_i^*({\bf r})(-\frac{1}{2} \nabla^2) \phi_i({\bf r})d^3 {\bf r}
 \nonumber\\
&\equiv& \int t_{\rm orb}({\bf r}) d^3 {\bf r}.
\label{A1}
\end{eqnarray}
The remainder, $T - T_{\rm s}$, is included in the exchange-correlation (XC)
functional $E_{\rm xc}[n]$.  Since successful 
$E_{\rm xc}$ approximations assume this KS KE decomposition, 
we focus on $T_{\rm s}$. This approach also evades the 
formidable task associated with the full $T[n]$ just mentioned.

For $T_{\rm s}[n]$ an explicit functional of $n$, the DFT total
energy functional is orbital-free:
\begin{eqnarray}
E^{\rm OF\mbox{-}DFT}[n]&=&T_{\rm s}[n]+E_{\rm Ne}[n]+E_{\rm H}[n] \nonumber \\ [8pt]
&&+\,E_{\rm xc}[n] + E_{\rm NN}.
\label{A2}
\end{eqnarray}
Here $E_{\rm Ne}[n]$ is the nuclear-electron interaction energy
functional, $E_{\rm H}[n]$ is the Hartree functional (classical
electron-electron repulsion), and $E_{\rm NN}$ is the inter-nuclear 
repulsion.  Then the variational principle gives the
single Euler equation
\begin{equation}
\frac{\delta T_{\rm s}[n]}{\delta n({\bf r})}+v_{\rm KS}([n];{\bf r})=\mu,
\label{A3}
\end{equation}
where $\mu$ is the Lagrange multiplier for density 
normalization $\int n({\bf r}) d^3 {\bf r} = N$ at the 
nuclear configuration ${\bf R}_{1}, {\bf R}_{2}, \ldots$, 
and 
$v_{\rm KS} = \delta (E_{\rm Ne} + E_{\rm H} + E_{\rm xc})/\delta n$.
The force on nucleus $I$ at ${\bf R}_{I}$ is  simply 
\begin{eqnarray}
{\bf F}_{I}&=& -\nabla_{{\mathbf R}_{I}} E^{\rm OF\mbox{-}DFT} %
\nonumber \\ [8pt] 
&=& -\nabla_{{\mathbf R}_{I}} E_{\rm NN} %
- \int  n({\bf r})  \, \nabla_{{\bf R}_{I}}  v_{\rm Ne}  d^3 {\bf r} %
\nonumber \\
&& -\int \BIG{[}{14pt}\frac{\delta T_{\rm s}[n]}{\delta n({\bf r})}
+v_{\rm KS}([n];{\bf r}) \BIG{]}{14pt}
\nabla_{{\bf R}_{I}}  n({\mathbf r})\, d^3 {\bf r}.
\label{A4}
\end{eqnarray}
The third term in Eq.~(\ref{A4}) shows that the biggest 
error in the calculated force will come from the gradient of the approximate
$T_{\rm s}[n]$ functional, because the kinetic energy is an order of
magnitude larger then the magnitude of $E_{\rm xc}$ (which also
must be approximated in practice). As will be discussed, when the
development of approximate KE functionals  focuses on forces,
it is convenient to use
Eq.~(\ref{A4}) with number density $n({\mathbf r})$ from a conventional 
KS calculation
(with a specific approximate $E_{\rm xc}$) as
input. 

In constructing approximate functionals it is quite
common to begin with the Thomas-Fermi functional \cite{Thomas,Fermi}, 
\bea
T_{\rm TF}[n] &=& c_0 \int n^{5/3}({\mathbf r})\, d^3{\mathbf r} %
\equiv \int t_0 ({\mathbf r})\, d^3{\mathbf r} \,,\nonumber \\
c_0 & = & \frac{3}{10} (3\pi^2)^{2/3} \, .
\label{TFKE}
\eea
By itself, the TF functional is not an acceptable approximation,
 because of, for
example, the Teller non-binding theorem \cite{Teller}.  
A more productive route for our purposes is to
decompose $T_{\rm s}[n]$ into the von Weizs\"acker energy $T_{\rm W}$
\cite{Weizsacker}, plus a \emph{non-negative} remainder, known as the
Pauli term $T_{\theta}$ \cite{TalBader78,BartolottiAcharya82,%
Harriman87,LevyOu-Yang88}, 
\begin{equation}
T_{\rm s}[n]=T_{\rm W}[n]+T_{\theta}[n], ~T_{\theta}[n] \; \geq 0   
\label{B7}
\end{equation}
with 
\begin{equation}
T_{\rm W}[n] = 
\frac{1}{8}\int  \frac{|\nabla n({\bf r})|^2}
 {n({\bf r})}  d^3{\bf r} \equiv \int  t_{\rm W} [n({\mathbf r})]  d^3{\bf r}
\,.
\label{B1a}
\end{equation}
Previously we have shown \cite{Perspectives}
 that the non-negativity of $T_\theta$
and $t_{\theta}({\bf r})$, defined by 
\bea
T_\theta&=&\int t_{\theta}({\bf r})d^3{\bf r} \nonumber \\
t_{\theta}& \equiv &t_{\rm orb}-\sqrt{n}\;\frac{1}{2} \nabla^2 \sqrt{n}
\label{B1b}
\eea
is a crucial but not sufficient condition for determinining a
realistic OF-KE approximation.  Details of some remaining issues follow.

\section{GGA-type KE functionals and their limitations}

\subsection{Basic structure}

Pursuit of local approximations for 
$t_{\rm orb}({\bf r}) = t_{\rm W}[n({\mathbf r}), \nabla n({\mathbf r})] %
+ t_\theta[n({\mathbf r}), \nabla n({\mathbf r}), \ldots]$ stimulates 
consideration of a counterpart to the 
generalized gradient approximation (GGA) for XC \cite{Perdew92}, namely
\begin{equation}
T^{\rm GGA}_{\rm s}[n]=c_0\int n^{5/3} ({\bf r}) F_{\rm t}(s({\bf r})) %
 d^3{\bf r} \, .
\label{B6}
\end{equation}
Here  $s$ is a dimensionless reduced density gradient
%
%
\be
s  \equiv \frac {|\nabla n|}{2nk_F} \; , \;\;\;\;\; %
k_F \equiv (3\pi^2 n)^{1/3} \,\,.
\label{sdefn}
\ee
$F_{\rm t}$ is a kinetic energy enhancement factor which goes to 
unity for uniform density. 
Equation (\ref{B6}) is motivated in part by the
conjointness conjecture \cite{LeeLeeParr91}, which  posits that 
$F_{\rm t} (s) \propto F_{\rm x}(s)$ where $F_{\rm x}$ is the 
enhancement factor in GGA exchange.
We showed previously that this relationship cannot hold 
strictly \cite{Perspectives}, but the form is suggestive and useful.

For connection with  $T_\theta \ge 0$, we
re-express  $T_{\rm W}$ in a form parallel  with Eq.~(\ref{B6}). From 
Eqs.\ (\ref{B1a}) and (\ref{sdefn}),
\begin{equation}
T_{\rm W}[n]=c_0\int n^{5/3} ({\bf r}) \frac{5}{3} s^2({\bf r}) d^3{\bf r} \; .
\label{B8}
\end{equation}
Then Eq.\ (\ref{B7}) gives 
\be
T^{\rm GGA}_s[n] = T_{\rm W}[n]+ c_0 \int n^{5/3}({\bf r}) 
F_{\theta}(s({\bf r})) d^3{\bf r} \, ,
\label{B10}
\ee
where
\begin{equation}
F_{\theta}(s) = F_{\rm t}(s) - \frac{5}{3}s^2 .
\label{Ft2}
\end{equation}
The final term of Eq.\ (\ref{B10}) thus is 
a formal representation of the GGA Pauli term $T^{\rm GGA}_{\theta}$.
Note that the form of Eq.\ (\ref{B10}) automatically
preserves proper  uniform scaling of $T_s$ (see Ref.~\onlinecite{Sham70}):
\bea
T_s[n_\gamma] & = & \gamma^2 \, T_s[n]\, , \nonumber \\
n_\gamma({\mathbf r}) & \equiv & \gamma^3 n(\gamma{\mathbf r})\, .
\label{Tunifscal}
\eea

Constraints that must be satisfied by the enhancement
factors associated with any satisfactory GGA KE functional include
%
\be
t_{\theta}([n];{\bf r})\; \geq 0 ,
\label{TthetaPos}
\ee
as well as \cite{LevyOu-Yang88,LevyPerdewSahni84,Herring86} 
\be
v_{\theta}([n];{\bf r})=\delta T_{\theta}[n]/\delta n({\bf r}) \geq 0\,, \;\;
 \forall \; {\mathbf r} \, .
\label{vthetaPos}
\ee
The quantity $v_\theta$ is known as the Pauli potential.  
Constraint Eq.\ (\ref{TthetaPos}) implies the non-negativity of 
the GGA enhancement
factor, $F_{\theta}(s({\bf r})) \geq 0 $.

For a slowly varying density that is not itself small, we have $s\approx 0$,
and it is appropriate to write $T_s$ as a gradient expansion \cite{Hodges}:

\begin{equation}
T_{\rm s}[n]=T_{\rm TF}[n]+\tfrac{1}{9}T_{\rm W}[n]+ \mbox{higher order terms}  .
\label{B2}
\end{equation}
Truncation at second order in $s$ gives the
second-order gradient approximation (SGA), with the
SGA enhancement factor \cite{Perdew92}
\begin{equation}
F_{\rm t}^{\rm SGA}(s) = 
1+\frac{1}{9}\cdot \frac{5}{3} s^2
=1+\frac{5}{27} s^2 \, , 
\label{B17}
\end{equation}
or 
\be
F_{\theta}^{\rm SGA}(s) = 1-\frac{40}{27} s^2.
\label{B18}
\ee
These forms should be exhibited by the  exact functional
in the limit of small density variation.
(Though there are $s \rightarrow \infty$ constraints
\cite{DreizlerGrossBook}, we have not used them so far.)
%

For GGA functionals, $v_{\theta}$,   
Eq.\ (\ref{vthetaPos}), can be written \cite{Gelfand-Fomin,Korn} as 
\be
\frac{\delta T_\theta^{\rm GGA}}{\delta n({\mathbf r})} %
 =   \frac{\partial t_\theta [n ({\mathbf r}), \nabla n ({\mathbf r})]} %
{\partial n({\mathbf r})} %
-  \nabla \, \cdot \, \frac{\partial t_\theta [n ({\mathbf r}), %
\nabla n ({\mathbf r})]}%
{\partial (\nabla n({\mathbf r}))} \,.
\label{GF}
\ee
After some tedium, one finds
\bea
v_\theta^{\rm GGA}& = & \frac{5}{3}c_0 n^{2/3}F_\theta + \nonumber \\
&& \!\!\! c_0n^{5/3} \frac{\partial F_\theta}{\partial s} %
 \left\lbrack  \frac{\partial s}{\partial n} %
-\frac{5}{3}\frac{\nabla n}{n}  \,\cdot\, %
\frac{\partial s}{\partial \nabla n} -  %
\nabla \,\cdot\, \frac{\partial s}{\partial \nabla n} \right \rbrack %
\nonumber \\
&& - c_0 n^{5/3}\frac{\partial^2 F_\theta}{\partial s^2} %
\left( \nabla s \,\cdot\,  \frac{\partial s}{\partial \nabla n} \right) 
\,.
\label{B21}
\eea
(The last line was omitted in Eq.\ (34) of Ref.\
\onlinecite{Perspectives} but included in the actual numerical work.) A 
somewhat cleaner expression that also makes it easier to understand
the extension we present below comes from shifting to the variable $s^2$ 
and defining both the reduced Laplacian density  $p$ 
\be
p \equiv \frac {\nabla^2 n}{(2k_F)^2 n} %
= \frac{\nabla^2 n}{4(3\pi^2)^{2/3}n^{5/3}} \; ,
\label{pdefn}
\ee
and one of the various possible dimensionless fourth-order derivatives $q$ 
\be
 q \equiv\frac{\nabla n \cdot (\nabla \nabla n) \cdot \nabla n}{(2k_F)^4 n^3}
=  \frac{\nabla n \cdot (\nabla \nabla n) \cdot \nabla n}
{16(3\pi^2)^{4/3}n^{13/3}} 
\label{qdefn}
\ee
(Note that our $s^2$, $p$ are denoted as $p$, $q$ respectively
in Ref.\ \onlinecite{Perdew07}.)
Then Eq.\ (\ref{B21}) becomes 
\bea
v_{\theta}^{\rm GGA}(s^2) &=&c_0 n^{2/3} %
\left\lbrace \frac{5}{3}F_\theta (s^2) %
- \left(\frac{2}{3}s^2 + 2p \right) \frac{\partial F_\theta}{\partial (s^2)} %
\right. \nonumber \\
&& \left. %
+ \left(\frac{16}{3} s^4 - 4q \right) \frac{\partial^2 F_\theta}{\partial %
 (s^2)^2} \right\rbrace \, .
\label{vthetas2pq}
\eea
See Appendix A for details.

\subsection{Singularities}

Near a point nucleus of charge $Z$ at the origin, 
the number density behaves to first order in $r$ as 
\be
n({\mathbf r}) \sim (1-2Z|{\mathbf r}|)+\rm O(|{\mathbf r}|^2)
\label{denscusp}
\ee
as required by Kato's cusp condition 
\cite{Kato57,Bingel63,PackByersBrown66,March..VanDoren2000,KryachkoLudena}.
Sufficiently close to a nucleus therefore, $n({\mathbf r})$
behaves as a Hydrogen-like 1s-electron density
\be
n_{H}({\mathbf r}) \sim \exp(-2Z |{\mathbf r}|)
\label{denscusp2}
\ee
That is, the variation of $n'(r)/n(r)$ is equal for these densities
sufficiently close to the nucleus, hence the form in 
Eq.\ (\ref{denscusp2}) is 
a reasonable near-nucleus approximation \cite{dens}. 
Consequences of differences in 
the higher-order terms in the respective Taylor expansions of
actual and hydrogenic 1s densities  
are discussed  in Section IV-A.  Also see Ref.\ \onlinecite{Janosfalvi05}
for a related discussion.

For $n({\bf r})$ of the form of Eq.~(\ref{denscusp2})
near $r=0$, $s$ and $q$
remain finite while $p \rightarrow -4Z/(2k_F)^2 r $.  In this
case, Eq.~(\ref{vthetas2pq}) becomes 
\begin{equation}
v_\theta^{\rm GGA}(r\rightarrow 0) = \frac{3Z}{5 r}
\frac{\partial F_\theta^{\rm GGA}}{\partial(s^2)}
+ \mbox{nonsingular terms.}
\label{B22}
\end{equation}
If $s^2$ is sufficiently small that it is a good approximation
to write $F_\theta^{\rm GGA} \approx 1+a s^2$ (note that this
is exactly the form of $F_\theta^{\rm SGA}$), Eq.~(\ref{B22})
simplifies to
\begin{equation}
v_\theta^{\rm GGA}(r\rightarrow 0) = \frac{3aZ}{5 r}
+ \mbox{nonsingular terms,}
\label{B22a}
\end{equation}
in which case $v_\theta^{\rm GGA}$ tends to infinity at the nuclei
with the same sign as the GGA parameter $a$.  The small values
of $s^2$ at the nuclei make this a general phenomenon.
(Near typical nuclei ($r \rightarrow 0$), numerical experience
shows that $s^2 \approx 0.15$, so that the small-$s^2$
behavior of any  $F_{\theta}(s^2)$ of GGA 
form  is relevant there.)

Equation (\ref{B22a}) shows that purely GGA Pauli potentials have
singularities in the vicinity of 
nuclear sites.  In
contrast, calculations using KS quantities as inputs show that the
exact Pauli potential is finite at the nuclei (see, for example, Ref.\
\onlinecite{BartolottiAcharya82} as well as Fig.\ \ref{SiO-tv_theta-R1} below).  
Moreover, the positivity requirement for $v_\theta$ will certainly be
violated near the nuclei both for $F_\theta^{\rm SGA}$ 
and for any GGA form with $a < 0$.

\vspace*{-12pt}
\subsection{Positivity: tests and enforcement}

To explore these positivity constraints, we tested six published KE
functionals
\cite{TranWesolowski02,PerdewBurkeErnzerhof96,AdamoBarone02,LacksGordon93,DePristoKress87,Thakkar92}
that either are strictly conjoint or  are based closely on conjointness.
The test used the diatomic molecule SiO, an important reference
species for us.  With LDA XC, we did a conventional,
orbital-dependent, KS calculation as a function of bond length (details
are in Appendix B).  At each bond length, the converged KS
density was used as input to the orbital-free $E[n]$ corresponding to 
one of the six $T_{\rm s}[n]$ approximations.  None predicted a stable SiO
molecule.  All six produced $a \le 0$ in Eq.\ (\ref{B22a}), hence all six
have non-trivial violations of $v_\theta$ positivity, with all the
effective enhancement factors very close to that of the SGA,
$F_{\theta}(s)=1-40/27s^2$.  Details are in Refs.\
\onlinecite {Perspectives} and \onlinecite{Signpost}.  Because of the constraint
 violation, conjointness thus can, at most, be a guide.  

We enforced positivity of $v^{\rm GGA}_\theta$ by 
particular parameterization  of $F_{\rm t}(s)$ 
forms based, in part, on the 
Perdew, Burke, and Ernzerhof (PBE) \cite{PerdewBurkeErnzerhof96} GGA XC
form: 
\bea
F_{\rm t}^{\rm PBE\nu}(s) &=& 1 + \sum_{i=1}^{\nu -1} C_i \left[\frac{s^2}{1+a_1 s^2}
     \right]^i , \; \nu = 2,3,4 \nonumber\\ [8pt]
F_{\rm t}^{\rm exp4}(s) &=& C_1(1-e^{-a_1s^2})+C_2(1-e^{-a_2 s^4}).
\label{PBEnuExp}
\eea
Our PBE2 form is the same as that used (with different parameters) by 
Tran and Wesolowski \cite{TranWesolowski02} while PBE3 corresponds to
the form introduced by Adamo and Barone
\cite{AdamoBarone02}, but, again, with different parameters.  
Quite similar forms also were explored by King and Handy \cite{KH0001}
in the context of directly fitting a KS kinetic potential
$v_s = \delta T_s / \delta n$ to conventional KS eigenvalues
and orbitals; see Eq.\ (\ref{D-1a}) below.

We fitted the parameters in the enhancement factors, Eqs.\ (\ref{PBEnuExp}),
to match the conventional KS internuclear
forces for various nuclear configurations of a three-molecule
training set: SiO, H$_4$SiO$_4$, and H$_6$Si$_2$O$_7$.  With
conventional KS densities as input, we found
semi-quantitative agreement with the conventional KS calculations for
single bond stretching in H$_4$SiO$_4$, H$_6$Si$_2$O$_7$, H$_2$O, CO,
and N$_2$.  All had energy minima within 5 to 20\% of the conventional
KS equilibrium bond-length values.  The latter two molecules provide
particular encouragement, since no data on C or N was included
in the parameterization. Details, including parameter values, 
are in Ref.\ \onlinecite{Perspectives}.

\subsection{Analysis of fitted functional behavior}

Despite this progress,  there is a problem.  Although the 
PBE$\nu$ and exp4 forms can give
Pauli potentials that are everywhere positive, yielding $a >0$ in
Eq.~(\ref{B22a}), they are singular at the nuclei, in contrast to the
negative singularities of previously published forms. For clarity
about the developments which follow, observe that these nuclear-site
singularities occur in $v_\theta$, hence are distinct from the
intrinsic nuclear-site singularity of the von Weizs\"acker potential,
which by operation on Eq.~(\ref{B1a}) can be shown to be
\be
v_{\rm W} \equiv \frac{\delta T_W[n]}{\delta n({\bf r})}
 = \frac{1}{8}\left\lbrack \frac{|\nabla n|^2}{n^2} %
- \frac {2\, \nabla^2 n}{n} \right\rbrack .
\label{vw}
\ee
The intrinsic singularity near a nucleus follows from Eq.\ (\ref{denscusp})
and has the form
\be
v_{\rm W} = -\frac{2Z}{r} \, .
\label{vwNearNuc}
\ee

\begin{figure}
\epsfxsize=8.4cm
\centerline{\epsffile{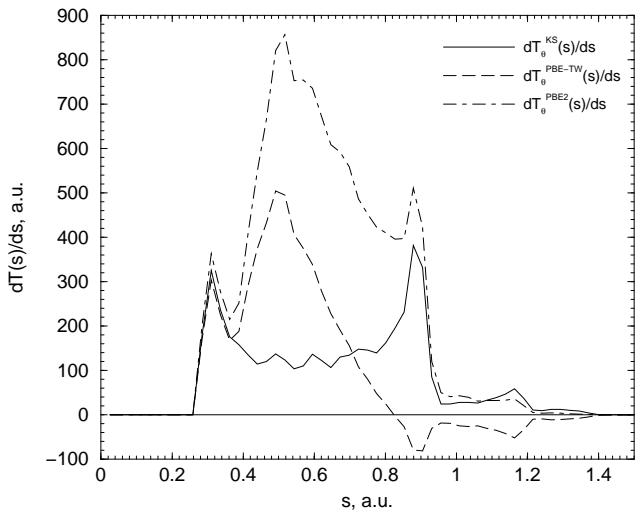}}
\caption{
 Energy density contributions to the Pauli term $T_\theta$
 as a function of $s$,
presented as values of $dT_\theta(s)/ds$ from Eq.~(\ref{D-1}); shown are
conventional Kohn-Sham $dT^{\rm KS}_\theta(s)/ds$ (the reference), our PBE2
functional, and the older PBE-TW GGA functional.
Data are for the SiO diatomic molecule at
bond length 1.926 {\AA} and are based on the density
from fully numerical KS-LDA computations.
}
\label{tss}
\end{figure}

Insight regarding the behavior of our modified conjoint functionals 
can be 
gained from consideration of the energy density
 $dT^{\rm appx}_\theta(s)/ds$ as a
function of $s$ for various functionals indicated by the 
generic superscript ``appx''.   This quantity  comes 
from differentiation of the
integrated contribution $T^{\rm appx}_\theta(s)$
 of the region $s({\bf r})\le s$ to the
kinetic energy: 
\begin{eqnarray}
T^{\rm appx}_\theta(s)& \equiv &
\int_{s({\bf r})\le s}t^{\rm appx}_\theta([n];{\bf r}) d^3{\bf r} \nonumber\\
&=& 
\int_0^s ds \int t^{\rm appx}_\theta([n];{\bf r})
 \delta(s-s({\bf r}))d^3{\bf r}\,. \hspace{16pt}
\label{D-1}
\end{eqnarray}
Figure \ref{tss} 
 shows $dT^{\rm appx}_\theta(s)/ds$  for the SiO molecule
at bond length $R=1.926$ \AA~(slightly stretched).  Values are shown 
%
for our recent parameterization PBE2 that
respects positivity, the Tran-Wesolowski parameterization
of the same form (PBE-TW), and for the exact, orbital-dependent 
KS Pauli term calculated from  \cite{LevyOu-Yang88}
\be
t^{\rm KS}_{\theta} %
= t_{\rm orb} -\left( \frac{1}{8}\frac{|\nabla n|^2}{n}-%
\frac{1}{4}\nabla^2 n \right)  \; , 
\label{D-1a}
\ee
where $t_\theta$  
and $t_{\rm orb}$ are defined in Eqs.\ (\ref{B1b}) and (\ref{A1})
respectively.  Recall that the exact value of $t_\theta$ must
be non-negative  \cite{LevyOu-Yang88}.  
For clarity,
note also that while $t_{\rm orb}$ can be negative, the equivalent form
\be
t_s \equiv t_{\rm orb} + \frac{1}{4}\nabla^2 n 
\label{tsubs}
\ee
is positive definite \cite{DreizlerGrossBook}. 

Figure \ref{tss} also shows that both approximate functionals closely 
resemble the exact KS kernel for 
$0.24<s<0.38$. But both of them have 
a much larger second peak around $s\approx 0.5$. In contrast,
the exact functional actually has a long low region before a 
second peak  at $s\approx 0.9$.
Our PBE2 approximate functional mimics the true second peak
via a too-strong third peak while the conventional GGA
PBE-TW functional has a spurious minimum at this point. Moreover, the PBE-TW
Pauli term goes negative for all $s>0.82$.  In addition, we see from
Fig.\ \ref{tss}  that 
the KS kinetic energy is nearly
totally determined by the behavior of $F_{\theta}$ over a relatively
small range of $s$, approximately $0.26\leq s \leq 1.30$ for the SiO diatomic.
The asymptotic regions 
($s\rightarrow 0$ and $s\rightarrow \infty$) do not play a significant role.
The range $0.26\leq s \leq 0.9$ has the highest weight of contribution 
(the highest differential contribution).
As an aside, we remark that PBE2 overestimates the KE 
presumably because it was fitted purely to forces 
without regard to total energies.

%
%
%
%
%
%
\begin{figure}
\begin{tabular}{c}
\epsfxsize=8.4cm
\epsffile{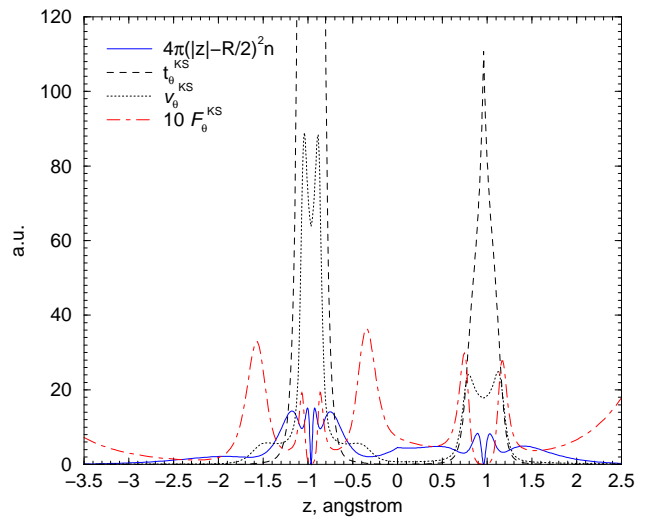}
\end{tabular}
\caption{
Conventional (reference) KS values for electronic density (scaled by the factor 
$4\pi(|z|-R/2)^2$, with $R$ the internuclear distance), 
Pauli term $t_{\theta}$, Pauli potential $v_{\theta}$, and
enhancement factor $F_{\theta}$,
 calculated for points on the
internuclear axis using KS LDA fully numerical orbitals for the SiO molecule;
Si at (0,\,0,\,$-0.963$){\AA}, O at (0,\,0,\,$+0.963$){\AA}.
}
\label{SiO-tv_theta-R1}
\end{figure}

\begin{figure}
\begin{tabular}{c}
\epsfxsize=8.4cm
\epsffile{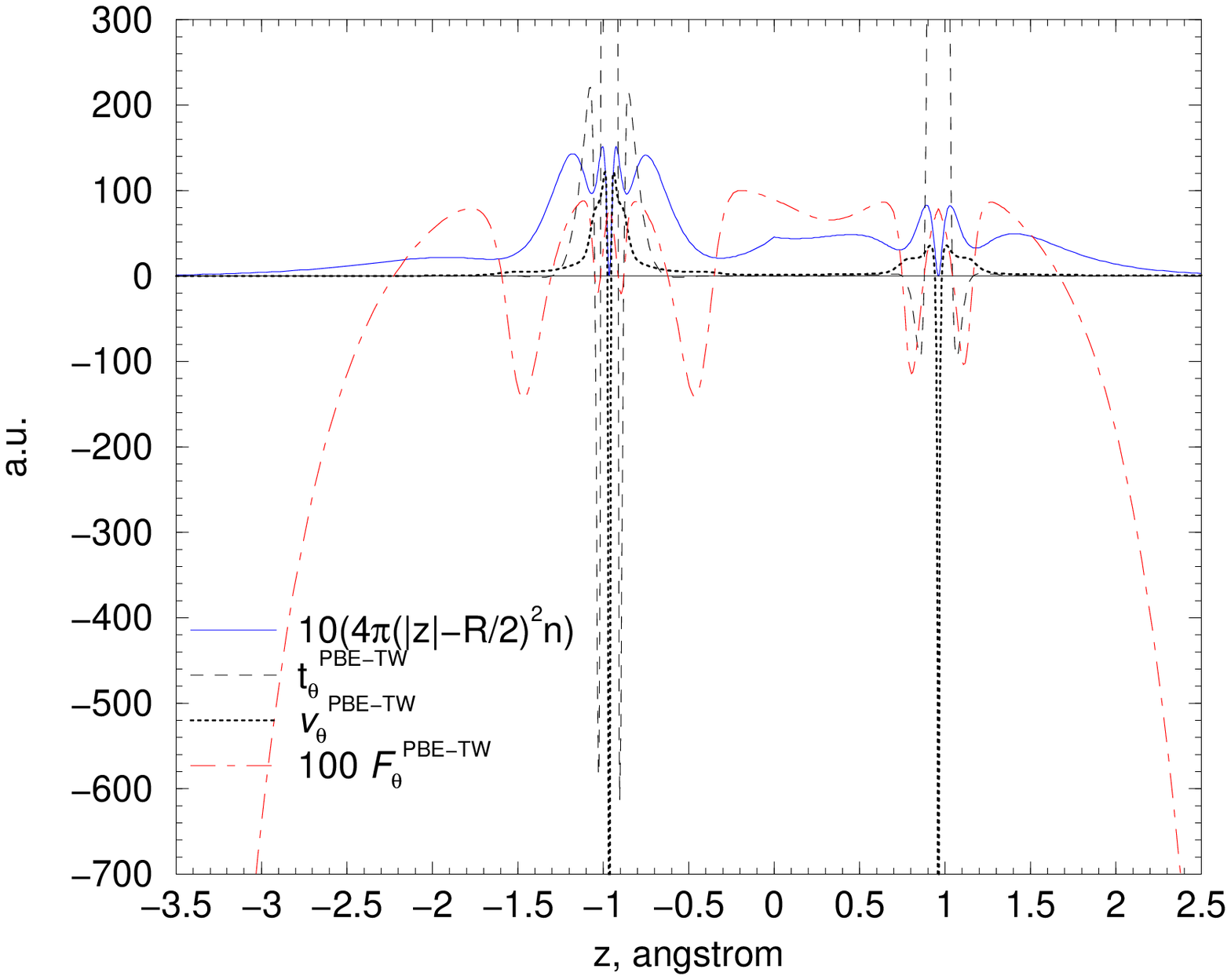}
\end{tabular}
\vskip -0.0cm
\begin{tabular}{c}
\hspace{-4.5cm} \epsfxsize=4.2cm \epsffile{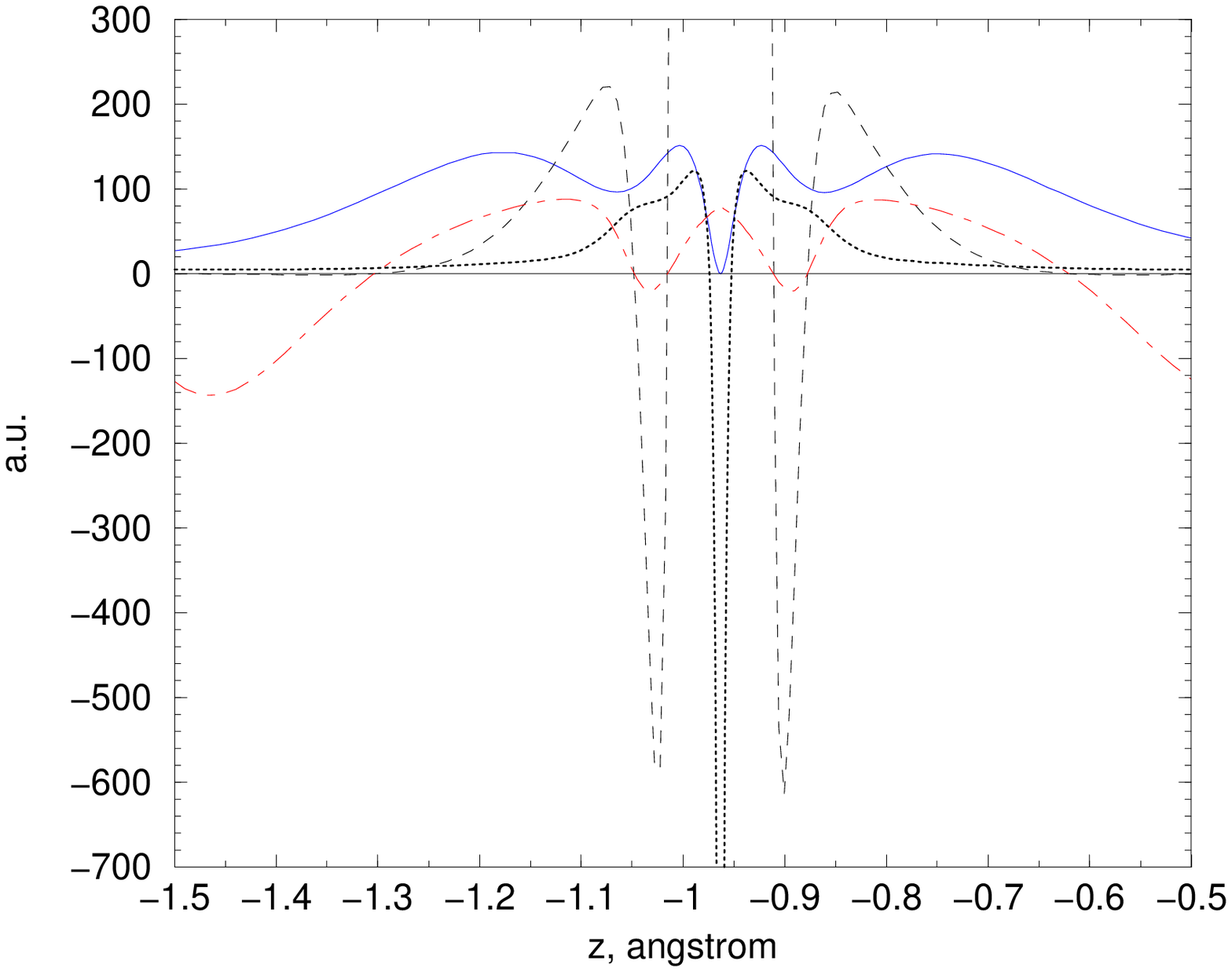}
\end{tabular}
\vskip -3.5cm
\begin{tabular}{c}
\hspace{4.0cm} \epsfxsize=4.2cm \epsffile{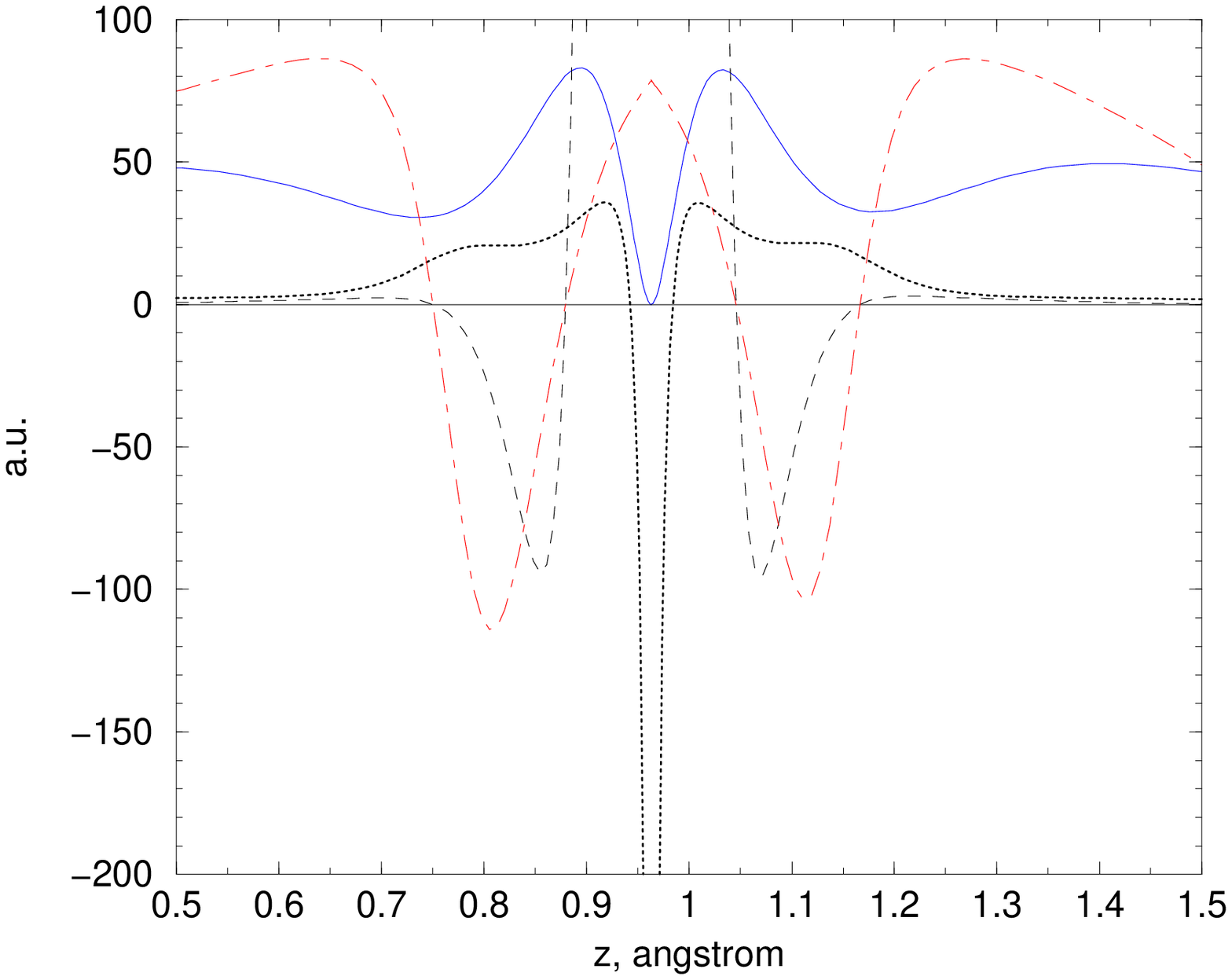}
\end{tabular}
\caption{
As in Fig. \ref{SiO-tv_theta-R1} for the PBE-TW conjoint approximation.
Lower left panel: region near Si-center; Lower right panel: region near 
O-center. 
}
\label{SiO-tv_theta-R2}
\end{figure}

\begin{figure}
\begin{tabular}{c}
\epsfxsize=8.4cm
\epsffile{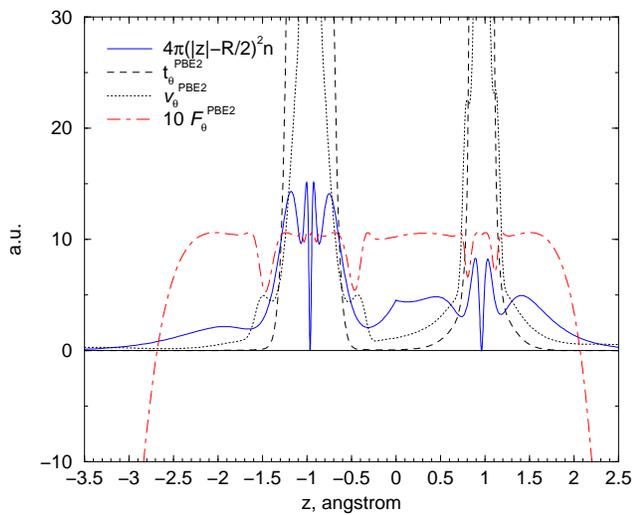}
\end{tabular}
\caption{
As in Fig. \ref{SiO-tv_theta-R1} for the
PBE2 modified-conjoint approximation. 
}
\label{SiO-tv_theta-R3}
\end{figure}

Figures \ref{SiO-tv_theta-R1}, \ref{SiO-tv_theta-R2}, and 
\ref{SiO-tv_theta-R3} 
provide comparisons of the reference $t_{\theta}^{\rm 
KS}$, $v_{\theta}^{\rm KS}$, and $F_{\theta}^{\rm KS}$ 
(where $F_{\theta}^{\rm KS}\equiv t_{\theta}^{\rm KS}/c_0 n^{5/3}$)
with the corresponding quantities for the PBE-TW and
PBE2 approximations.  The values 
are along the internuclear axis of the SiO
molecule with internuclear separation 1.926 {\AA}.  
The KS Pauli potential was calculated using the exact orbital-dependent
expression
\be
v_{\theta}([n];{\bf r})=\frac{t_{\theta}([n];{\bf r})}{n({\bf r})} 
+
\sum_{i=1}^{N}  %
(\varepsilon_{N}-\varepsilon_i)\frac{|\phi_i({\bf r})|^2}{n({\bf r})}
\, ,
\label{VthKS}
\ee
where $\phi_i$ and $\varepsilon_i$ are the occupied KS orbitals and
eigenvalues respectively. Equation (\ref{VthKS}) is obtained in a way
similar to that used in Ref.\ \onlinecite{LevyOu-Yang88};
 see Ref.\ \onlinecite{Signpost}.

In Fig.\ \ref{SiO-tv_theta-R1}, all three KS quantities,
$t_{\theta}^{\rm KS}$, $v_{\theta}^{\rm KS}$, and $F_{\theta}^{\rm
KS}$ are everywhere {\it non-negative}, as they must be.  Observe
that  $v_{\theta}^{\rm KS}$ is {\it finite} at the nuclei and
has local maxima in positions close to the inter-shell minima of the
electronic density.  

In contrast, the energy density of the PBE-TW Pauli term and
corresponding enhancement factor have negative peaks in the
inter-shell regions, violations of the non-negativity constraint for
$t_{\theta}$.  However, addition to $t_{\theta}^{\rm appx}$
 of any multiple of a Laplacian term $\nabla^2 n$
 would change only the local
behavior without altering the value of $T^{\rm appx}_{\theta}[n]$,
 so the PBE-TW
misbehavior might be resolved by such an addition. See discussion
below. Figure \ref{SiO-tv_theta-R2} also shows that $v_\theta$ for the
PBE-TW functional has very sharp negative peaks exactly at the nuclear
positions, in accord with Eq.\ (\ref{B22a}).

Figure \ref{SiO-tv_theta-R3}  shows that the 
modified conjoint
PBE2 $v_\theta$  respects the
positivity constraint everywhere.  Inter-nuclear forces in the
attractive region are described at least qualitatively correctly as a
result.  The PBE2 potential
is still divergent at the nuclei in accordance with Eq.\
(\ref{B22a}).

As an aside we remark on two small computational issues.
First,  the absence of nuclear-site
singularities in the computed correct Pauli potentials
$v_{\theta}^{\rm KS}$ might be argued to occur because the computed
density does not have precisely the proper nuclear site cusp, i.e., does
not strictly obey Eq.\ (\ref{denscusp}).  However, our numerical results
are consistent with those in Ref.\ \onlinecite{BartolottiAcharya82}. Those
authors used numerical orbitals \cite{BartolottiPrivCommun} which 
presumably satisfied the cusp condition approximately.  More
importantly, if  
a specific numerical technique gives a
KS density and associated KS $t_s$  that produce a non-singular $v_\theta$, 
then an approximate $v_\theta$ 
evaluated with the \emph{same} density should not introduce singularities.
Second, the reader may notice that Fig.~\ref{SiO-tv_theta-R3} shows small
negative values for $t_\theta$ far from the bonding region of the
molecule.  This behavior is due to numerical imprecision associated
with computation for extremely small values of the density, and makes
no appreciable contribution to the kinetic energy.

Finally, an insight to the harm of excess positivity of $v_\theta$ 
can be seen by examining the dependence of $T_\theta[n]$ upon $v_\theta$.
 From a known virial relation  \cite{LevyOu-Yang88} we have 
\begin{equation}
T_{\theta}[n]= \frac{1}{2 }\int v_{\theta}([n];{\bf r})
(3+{\bf r}\cdot \nabla)n ({\bf r})d^3{\bf r}.
\label{D1}
\end{equation}
Any spurious singularities of $v_\theta^{\rm appx}$ at the nuclei
clearly will cause special problems in overweighting the integrand. 

\section{Beyond GGA-type Functionals} 

The preceding analysis makes clear the need for 
more flexible functionals than
the forms  of Eq.\ (\ref{PBEnuExp}).  
In particular, 
nuclear site divergences of $v_\theta$ 
are  unavoidable for all purely GGA-type functionals
{\it e.g.}, GGA, GGA-conjoint, and modified conjoint KE functionals;
recall  Eqs.\ (\ref{B22}) and (\ref{B22a}).  Additional
variables and constraints upon them are required to eliminate the
singularities.

\subsection{Reduced derivatives of the density}

Consider again the gradient expansion
of $T_{\rm s}[n]$, Eq.\ (\ref{B2}) 
(see Refs.~\onlinecite{Hodges}, \onlinecite{Murphy81,Yang86,Yang..Lee86} for details),
which we recast as
\be
T_{\rm s}[n]=\int \Big\{t_{0}([n];{\bf r})+t_{2}([n];{\bf r})+t_{4}([n];{\bf r})+...\Big\}d^3{\bf r}\, .
\label{II10}
\ee
Here $t_0$ is as in Eq.\ (\ref{TFKE}), $t_{2} = (1/9)\, t_{\rm W}$, and
\bea
t_4([n];{\bf r})&=&\frac{1}{540(3\pi^2)^{2/3}}
n^{5/3}({\bf r})\BIG{[}{14pt}\Big(\frac{\nabla^2 n({\bf r})}{n^{5/3}({\bf r})}\Big)^2
\nonumber \\
&& \!\!\!\!\!\!\!\!\!\!\!\!\!\! %
-\frac{9}{8}\Big(\frac{\nabla n({\bf r})}{n^{4/3}({\bf r})}\Big)^2
\Big(\frac{\nabla^2 n({\bf r})}{n^{5/3}({\bf r})}\Big)
+\frac{1}{3}\Big(\frac{\nabla n({\bf r})}{n^{4/3}({\bf r})}\Big)^4\BIG{]}{14pt}\,.
\label{II11}
\eea
The sixth-order term, dependent upon $n$, $|\nabla n|$, $\nabla^2 n$,
$|\nabla\nabla^2 n|$, and $\nabla^4 n$, is given in Ref.~\onlinecite{Murphy81}.

As is well-known, in a finite system (e.g. molecule), a
Laplacian-dependent term $\nabla^2n$ affects only the local behavior
of the kinetic energy density.  Arguments have been advanced for and
against including such Laplacian dependence in the KE density
functional;
 for example, see Refs.~\onlinecite{KH0001}, \onlinecite{Yang..Lee86},
 \onlinecite{Ayers..Nagy01}.
Recently Perdew and Constantin \cite{Perdew07} presented a KE
functional that depends on $\nabla^2 n$ via a modified fourth-order
gradient expansion.  Though not stated that way, their functional
obeys the decomposition of Eq.\ (\ref{B7}).  It is intended to be
universal (or at least very broadly applicable), whereas we are
focused on simpler functionals that require parameterization to
families of systems. The Perdew-Constantin form involves a rather
complicated functional interpolation between the gradient expansion
and the von Weizs{\"a}cker functional. They did not discuss the
corresponding potential $v_\theta$ nor Born-Oppenheimer forces. And
they characterized the performance of their functional for the
energetics of small molecule dissociation as ``still not accurate
enough for chemical applications''.  So we proceed rather differently.

Rearrange the foregoing 
gradient expansion into $T_{\rm W} + T_\theta$ form (recall Eq.\
(\ref{B7})):
\bea
T_\theta[n] &\!\!=&\!\! \int\! \left\lbrack c_0 n^{5/3}({\mathbf r}) %
\left( 1 -\frac{40}{27} s^2\right) + t_4 + t_6 + \cdots \right\rbrack %
d^3{\bf r} \nonumber \\
&\!\!=&\!\! \int\! \left\lbrack t_0 \left( 1 -\frac{5}{3} s^2\right) %
+ t_0\frac{5}{27}s^2  + t_4 + t_6 + \cdots \right\rbrack %
d^3{\bf r} \nonumber \\
&\!\!\equiv&\!\! \int\! \left\lbrack t_{\theta}^{(0)}([n];{\bf r}) %
+t_{\theta}^{(2)}([n];{\bf r}) %
+t_{\theta}^{(4)}([n];{\bf r})+...\right\rbrack d^3{\bf r} %
\nonumber\\
\label{TthetaGradExpan}
\eea
where
\be
t_{\theta}^{(0)}([n];{\bf r})=t_{0}([n];{\bf r})%
\Big[1-\frac{5}{3}s^2\Big] \,,
\label{II13}
\ee
\be
t_{\theta}^{(2)}([n];{\bf r})=t_{0}([n];{\bf r})%
\Big[\frac{5}{27}s^2\Big] \,,
\label{II14}
\ee
and 
\be
t_{\theta}^{(4)}([n];{\bf r})=t_{0}([n];{\bf r})%
\Big[\frac{8}{81}\Big(p^2-\frac{9}{8}s^2p+\frac{1}{3}s^4\Big)\Big] \,.
\label{II15}
\ee
Each term, Eq.\ (\ref{II13})--(\ref{II15}), of 
Eq.\ (\ref{TthetaGradExpan}) can be put straightforwardly into a 
GGA-like form:
\be
t_{\theta}^{(2i)}([n];{\bf r})=t_{0}([n];{\bf r})F_{\theta}^{(2i)}(s,p,...)\,, 
\label{V1}
\ee
where $s$, $p$ are as in Eqs.\ (\ref{sdefn}) and (\ref{pdefn}) 
respectively. 
The first two terms of the expansion yield the SGA enhancement factor
already discussed
\be
F_{\theta}^{\rm SGA}\equiv F_{\theta}^{(0)}+F_{\theta}^{(2)}=1+a_2s^2\,,
\label{V2}
\ee
with $a_2=-40/27$.
The fourth-order (in highest power of $s$) term is
\be
F_{\theta}^{(4)}=a_4s^4+b_2p^2+c_{21}s^2p\,,
\label{V3}
\ee
with coefficients $a_4=8/243$, $b_2=8/81$, and $c_{21}=-1/9$.

Rather than retain those values of $a_2$, $a_4$, $b_2$, $c_{21}$,
we instead treat them as parameters and seek values or 
relationships among them which would yield a non-singular $v_\theta$
through a given order. (Corresponding improvement of Thomas-Fermi theory
by imposition of the nuclear cusp condition was introduced
in Ref.\ \onlinecite{ParrGhosh86}.)

Functional differentiation of each term in Eq.~(\ref{TthetaGradExpan})
gives the formal gradient expansion 
$v_{\theta}=v_{\theta}^{(0)}+v_{\theta}^{(2)}+v_{\theta}^{(4)}+ \cdots$,
where 
$F_\theta^{(2i)}$ (shown below with its arguments suppressed for clarity) 
is a function of $s^2$, $p$, and in principle, higher derivatives of 
$n({\bf r})$:
\begin{eqnarray}
v_\theta^{(2i)}({\bf r}) &=& t_0([n];{\bf r})\BIG{[}{14pt}
\frac{5}{3n({\bf r})}F_\theta^{(2i)}+\frac{\partial F_\theta^{(2i)}}
{\partial (s^2)}\frac{\partial (s^2)}{\partial n({\bf r})} \nonumber \\ [8pt]
&& +\,\frac{\partial F_\theta^{(2i)}}
{\partial p}\frac{\partial p}{\partial n({\bf r})}+\dots\BIG{]}{14pt}
\nonumber \\ [8pt]
&& -\,\nabla \cdot \left( t_0([n];{\bf r})\frac{\partial F_\theta^{(2i)}}
{\partial (s^2)}\frac{\partial (s^2)}{\partial \nabla n({\bf r})}\right)
\nonumber \\ [8pt]
&&+\,\nabla^2\left( t_0([n];{\bf r})\frac{\partial F_\theta^{(2i)}}
{\partial p}\frac{\partial p}{\partial \nabla^2 n({\bf r})}\right) + \cdots
 \,.\hspace{18pt}
\label{V4}
\end{eqnarray}
The ellipses in Eq.~(\ref{V4}) correspond to additional terms that are
needed only if
$F_\theta^{(2i)}$ depends upon derivatives other than $s$ and $p$.

After manipulation (see Appendix A), one obtains the potentials
corresponding to the enhancement factors in Eqs.~(\ref{V2}) and (\ref{V3}):
\bea
 v_\theta^{\rm SGA} &=& c_0 n^{2/3}\left[\frac{5}{3}+a_2 s^2
-2 a_2 p \right] \, ,  \label{V4a}\\ [8pt]
v_\theta^{(4)} &=& c_0n^{2/3}\BIG{[}{16pt}\left(11a_4+\frac{88}{9}\,c_{21}\right)
s^4 \nonumber \\
&& - \left(5b_2+2c_{21}\right)p^2 - \left(4a_4-\frac{80}{9}\,b_2\right)s^2p
\nonumber \\ [8pt]
&&-\,\left(8a_4+\frac{32}{3}\,c_{21}\right)q \nonumber \\
&& -\frac{20}{3}\,b_2 q^\prime
+2b_2 q^{\prime\prime}+2c_{21}q^{\prime\prime\prime}\BIG{]}{16pt} \, .
\label{V4b}
\eea
Here $q$ is as in Eq.\ (\ref{qdefn}) and 
$q^\prime$, $q^{\prime\prime}$, and $q^{\prime\prime\prime}$ 
are other dimensionless fourth-order reduced density derivatives defined
as
\begin{eqnarray}
 q^\prime&\equiv&\frac{\nabla n \cdot \nabla \nabla^2 n}
{(2k_F)^4n^2} =\frac{\nabla n \cdot \nabla \nabla^2 n}
{16(3\pi^2)^{4/3} n^{10/3}} \, ,  \\ [8pt]
 q^{\prime\prime}&\equiv&\frac{\nabla^4 n}{(2k_F)^4 n}= \frac{\nabla^4 n}
{16(3\pi^2)^{4/3} n^{7/3} } \, , \\ [8pt]
 q^{\prime\prime\prime}&\equiv&\frac{\nabla\nabla n \,:\,\nabla\nabla n}
{(2k_F)^4 n^2} = \frac{\nabla\nabla n \,:\,\nabla\nabla n}
{16(3\pi^2)^{4/3}n^{10/3} }  \, .
\end{eqnarray}
The operation denoted by the colon in the numerators of 
$q^{\prime\prime\prime}$  is $A:B \equiv \sum_{ij} A_{ij} B_{ji}$.

At Eq.~(\ref{B22}) we have already pointed out that an enhancement
factor of SGA form, specifically, that of Eq.~(\ref{V2}), produces
a Pauli potential
\be
v_{\theta}^{\rm SGA}(r) = v_{\theta}^{(0)}(r)+v_{\theta}^{(2)}(r)
=\frac{3}{5} \frac{Za_2}{r} +  \mbox{nonsingular terms.}
\label{V6}
\ee
This exhibits the $1/r$  SGA Pauli potential nuclear singularity 
already discussed; we return to this point in a moment. 

For the fourth-order enhancement factor, Eq.~(\ref{V3}),
we again note that $s$ and $q$ are non-singular
near the nucleus, while 
\be
\lim_{r \rightarrow 0} s^2(r) = Z^2/[3\pi^2n(0)]^{2/3} \;.
\label{s2lim}
\ee
With a density of the form of Eq.\ (\ref{denscusp2}), 
Eq.\ (\ref{V4b})  thus gives
 the near-nucleus behavior of the fourth-order potential as
\bea
v_{\theta}^{(4)}({r})
 &=& \frac{c_0}{16[9\pi^4n(r)]^{2/3}}
\left[-\frac{16Z^2}{3r^2}
\left(5b_2+3 c_{21}\right) \right. \nonumber \\ 
&& \left. +\frac{32Z^3}{9r}\left(18a_4 + 17b_2+18c_{21}\right)\right] %
\nonumber \\
&&  +\, \mbox{nonsingular terms.}
\label{V7}
\eea
The singularities in  $1/r^2$ and $ 1/r$ 
can be removed by requiring that the numerators of the
first two terms of\ Eq. (\ref{V7}) both vanish, or equivalently
\bea
c_{21}&=&-\frac{5}{3}b_2 \,, \nonumber\\
a_4&=&\frac{13}{18}b_2\,.
\label{V8}
\eea

In the spirit of the GGA, 
we are led to defining a fourth-order reduced density derivative (RDD) as 
\be
\kappa_4=s^4+\frac{18}{13}p^2-\frac{30}{13}s^2p \,. 
\label{V9}
\ee
This RDD with Eq.\ (\ref{V3}) gives an enhancement factor
\be
F_{\theta}^{(4)}(\kappa_4)=a_4\kappa_4 \,,
\label{Ftheta4reduced1}
\ee
which yields a Pauli potential with {\it finite} values at point nuclei.
Clearly it is not the only $\kappa_4$-dependent enhancement
factor  with that property.  So, we 
seek $F_\theta^{(4)}(\kappa_4)$ functional forms which are more general
than Eq.\ (\ref{Ftheta4reduced1}) and which give a positive-definite, non-singular 
$v_\theta$.

At this point, it is prudent to consider how many terms in the 
Taylor series expansion of the 
density Eq. (\ref{denscusp2}) are relevant for the cancellation of 
singularities in Eq.\ (\ref{V7}). The answer is four terms: 
$n(r) \propto 1 - 2Zr +2Z^2r^2 - (4/3)Z^3r^3$. That is, 
the singularities will 
reappear for a density of the form of  Eq.\ (\ref{denscusp}) if
the second- and third-order terms differ 
from those defined by a Hydrogen-like density expansion, e.g., 
Eq.\ (\ref{denscusp2}). Thus, the foregoing cancellation fails for a density
with power series expansion
$n(r) \propto 1 - 2Zr  - (4/3)Z^3r^3 + \ldots$.  This fact will 
limit applications of simple $\kappa_4$-based KE functionals to those 
densities which have precisely Hydrogen-like behavior up to fourth order.

It is necessary, therefore, to consider other candidates for RDD 
variables which 
would provide cancellation of singularities for the density 
Eq.\ (\ref{denscusp}) independently of hydrogenic higher-order terms 
in the Taylor series expansion of the density.
The observation that 
$\kappa_4\sim \rm O(\nabla^4)$ suggests that the effective
or operational order of $\nabla$ in such a candidate variable 
should be reduced to second order.  This in turn suggests 
a candidate variable, still based on the fourth-order gradient expansion 
Eq. (\ref{II15}), namely 
\be
F_{\theta}^{(4-2)}=\sqrt{a_4s^4+b_2p^2+c_{21}s^2p}\, ,
\label{V9a}
\ee
(compare Eq.\ (\ref{V3})). Now consider a density of the form
of  Eq.\ (\ref{denscusp}) 
but with arbitrary first- and higher-order near-nucleus 
expansion coefficients,
\be
n(r)\sim (1+C_1r+ C_2 r^2+C_3 r^3).
\label{V9bb}
\ee
Following the same lines as those used to reach Eq.\ (\ref{V7}), one finds
\be
v_{\theta}^{(4-2)}({r})\sim \frac{c_{21}}{\sqrt{b_2}} \frac{1}{r}
+ \mbox{nonsingular terms} \; .
\label{V7a}
\ee
The singular term would be eliminated by the choice $c_{21}=0$. 
The cancellation  is universal in that it does not depend on the density
expansion coefficients, $C_i$ (while the singular term prefactor 
and non-singular terms do, of course, depend on those expansion 
coefficients).  Hence a candidate RDD 
variable (denoted as $\tilde \kappa_4$)
which provides cancellation of singular terms in the Pauli potential
could be defined as
\be
\tilde \kappa_{4}=\sqrt{s^4+b_2 p^2} \;,\qquad b_2>0 \,. 
\label{V9b}
\ee
Note that this form is manifestly positive.

This RDD can be used to construct a variety of 
enhancement factors to replace Eq.\ (\ref{V3}) for the 
fourth-order approximation to the Pauli term, for example
$F_{\theta}(\tilde \kappa_4)=a_4\tilde\kappa_4$.
This simplest enhancement factor corresponds to 
a Pauli potential with {\it finite} values at point nuclei but clearly
it is not the only $\tilde\kappa_4$-dependent one with that property.
Any linear combination of non-singular enhancement factors 
(including the simple $F_\theta = 1$) also will be non-singular.
A combination of two PBE-like forms 
(see Eq.\ (\ref{VI2}) below)
%
%
is also non-singular, as can be 
checked analytically for any density with near-nucleus behavior 
defined by Eq.\ (\ref{V9bb}), hence also Eqs.\ (\ref{denscusp}), 
(\ref{denscusp2}).

There are, of course, $\tilde\kappa_4$-dependent functionals that yield a
divergent potential, e.g. 
\be
F_{\theta}(\tilde\kappa_4)= \tilde\kappa_4^2 
\label{Ftheta4reducedDiver}
\ee
%
so one must be cautious.  

Regarding the second-order forms, 
Eq.\ (\ref{V6}) shows that, short of complete removal
of the $s^2$ term from $F_\theta^{\rm SGA}$,  we cannot 
cure the singularity in $v_\theta^{\rm SGA}$. There is no direct 
analogy to the removal of singularities in $v_\theta^{(4)}$ just discussed.  
Instead, in parallel with Eq.\ (\ref{V9}) or Eq.\ (\ref{V9b}), 
we introduce a second-order RDD
\be  
\kappa_2=s^2+b_1p \, ,
\label{V10}
\ee
with $b_1$ to be determined.  Then, in analogy with a PBE-type
enhancement factor, we can define an enhancement factor dependent
only on second-order variables as 
\be  
F_{\theta}^{(2)}(\kappa_2)=\frac{\kappa_2}{1+\alpha \kappa_2} \,.
\label{V11}  
\ee
For it, the near-nucleus (small $r$) behavior of the Pauli potential 
is 
\be
v_{\theta}^{(2)}(r)= C_1^{(2)}
\frac{(1+C_2^{(2)}b_1\alpha)}{b_1\alpha^2}+ {\rm O}(r)\,,
\label{V12}
\ee
with constants $C_i^{(2)}>0$  which depend on the specific density
behavior being handled.

The RDDs considered thus far are
combinations of powers of $s$ and $p$ 
which ensure cancellation of nuclear cusp divergences in $v_\theta$.  
Thus, we define a class of approximate KE functionals, the
reduced derivative approximation (RDA) functionals, as 
those with enhancement factors depending on the RDDs
\be
T_{\rm s}^{\rm RDA}[n]\equiv T_{\rm W}[n]+\int t_0([n];{\bf r}) 
F_{\theta}(\kappa_2({\bf r}),\tilde\kappa_4({\bf r}))~ d^3 {\bf r} \,,
\label{V14}
\ee
($t_0[n]$ from Eq.\ (\ref{TFKE})) with non-divergent Pauli 
potentials as a consequence of 
constraints imposed on the coefficients in the RDDs. This route of
development of KE functionals is under active investigation; see below.

For insight, Figure \ref{kappas} shows the behavior of the RDD
$\tilde\kappa_4$ along the SiO internuclear axis for four values of
$b_2$.  The behavior in the vicinity of the Si atom is shown.  Both
the $s$ and $p$ variables have four maxima which lie close to
the intershell minima in the density.  Increasing the value of 
$b_2$ increases the height of the corresponding  maxima
for $\tilde\kappa_4$ RDD (because the contributions from the $p$-maxima
increase).

One of the peculiarities is that the reduced density Laplacian $p$ is
divergent at the nucleus and, as a consequence, $\tilde\kappa_4$ itself 
also is divergent, even though it generates a non-divergent $v_\theta$. 
This divergence will not affect the KE enhancement factors provided that
$\lim_{\tilde\kappa_4\rightarrow \infty} F_{\theta}(\tilde\kappa_4)=\rm %
 Constant$.
One of the advantages of the $\tilde\kappa_4$ variable is its positiveness 
everywhere  (by definition). 

\begin{figure}        
\epsfxsize=7.4cm
\vspace{0.50cm}
\epsffile{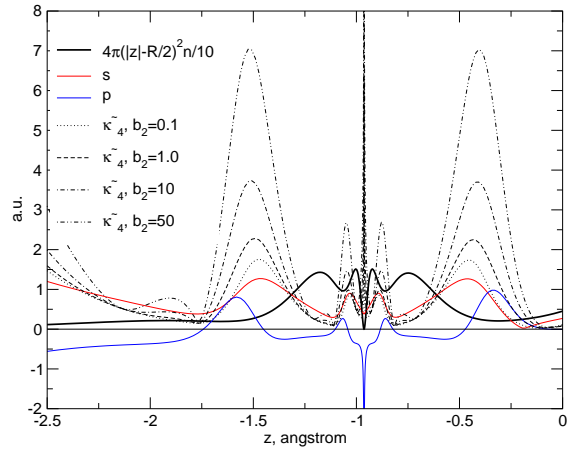}
\vspace{0.0cm}
\caption{
The fourth order, $\tilde\kappa_4$
reduced density derivative for different values of $b_2$ 
along the internuclear axis $z$
for the SiO
diatomic molecule near the Si atom: 
Si at (0,\,0,\,$-0.963$){\AA},
O at (0,\,0,\,$+0.963$){\AA}. 
Variables $s$ and $p$, 
are shown for comparison.
}
\label{kappas}
\end{figure}

\subsection{Parameterization of a RDA functional}

In addition to the positive singularities, another limitation of 
our earlier modified-conjoint type KE
functionals was the inability to parameterize them to provide both
forces and total energies simultaneously \cite{Perspectives}.  Given
the emphasis on MD simulations, 
parameterization to the forces was the priority.  With the
spurious repulsive singularities removed from RDD functionals,
the question arises whether total energy parameterization can be used
and, if so, if it is beneficial. The usual  {\it energy} fitting 
criterion is to minimize
\be
\omega_E=\sum_{i=1}^m \left|E^{\rm KS}_i-
E^{\rm OF\mbox{-}DFT}_i\right|^{\,2}\,,
\label{IV1}
\ee
over systems (e.g., atoms, molecules) and configurations (e.g., diatomic 
molecule bond length)
indexed generically here by $i$.  When the parameter adjustment is
done for fixed-density inputs (i.e., conventional KS densities as
inputs), this total energy optimization is equivalent to optimization
of the $T_{\rm s}$ functional.  We did
this for determination of the empirical parameters for the 
new RDA-type functionals $F_{\theta}^{\rm
RDA}=F_{\theta}(\tilde\kappa_4)$.

Since $F_\theta = 1$ (or any constant in general) 
also yields a non-singular Pauli potential, we
can form $\tilde\kappa_4$-dependent enhancement factors which resemble
GGA forms and thereby enable connection with the modified conjoint
GGA functionals discussed already.  One form which we have begun 
exploring (see below) is 
\bea
F_{\theta}^{{\rm RDA}(ij)}(\tilde\kappa_4)&=&A_0+
A_1 \left( \frac{\tilde\kappa_4}{1+ \beta_1\tilde\kappa_4}\right)^i %
\nonumber \\
&& +A_2 \left( \frac{\tilde\kappa_4}{1+ \beta_2\tilde\kappa_4}\right)^j %
\,.
\label{VI2}
\eea
$A_i$ and $\beta_i$ are parameters to be determined. 
Even this simple form has two desirable properties: (i) the
corresponding $v_\theta$ is finite for densities with the near-nucleus
behavior defined by Eq.\ (\ref{V9bb}), hence also  Eqs.\ (\ref{denscusp})
or \ (\ref{denscusp2}) (this has been checked by explicit analytical 
calculation); 
(ii) the divergence of
$\tilde\kappa_4$ near the nucleus (see Figure \ref{kappas}) cancels in
Eq.~(\ref{VI2}) ($\lim_{\tilde\kappa_4\rightarrow \infty}~F^{{\rm
RDA}(ij)}_{\theta}(\tilde\kappa_4)=A_0+A_1/\beta_1^i+A_2/\beta_2^j$). 
Positivity of $F^{{\rm RDA}(ij)}_{\theta}$ depends on 
the parameters $A_i$ and must be checked for any given determination 
of their values.

\begin{figure}                 
\epsfxsize=7.4cm
\vspace{0.5cm}
\epsffile{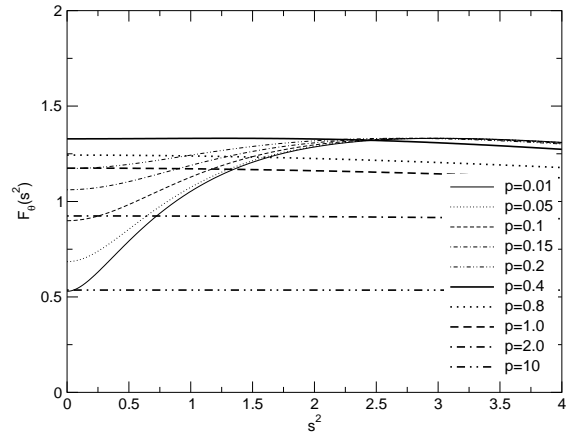}
\vspace{0.3cm}
\caption{
The RDA(24) enhancement factor as a function of $s^2$ for selected values 
of $p$.
}
\label{FtS2}
\end{figure}

\begin{figure}
\epsfxsize=8.4cm
\epsffile{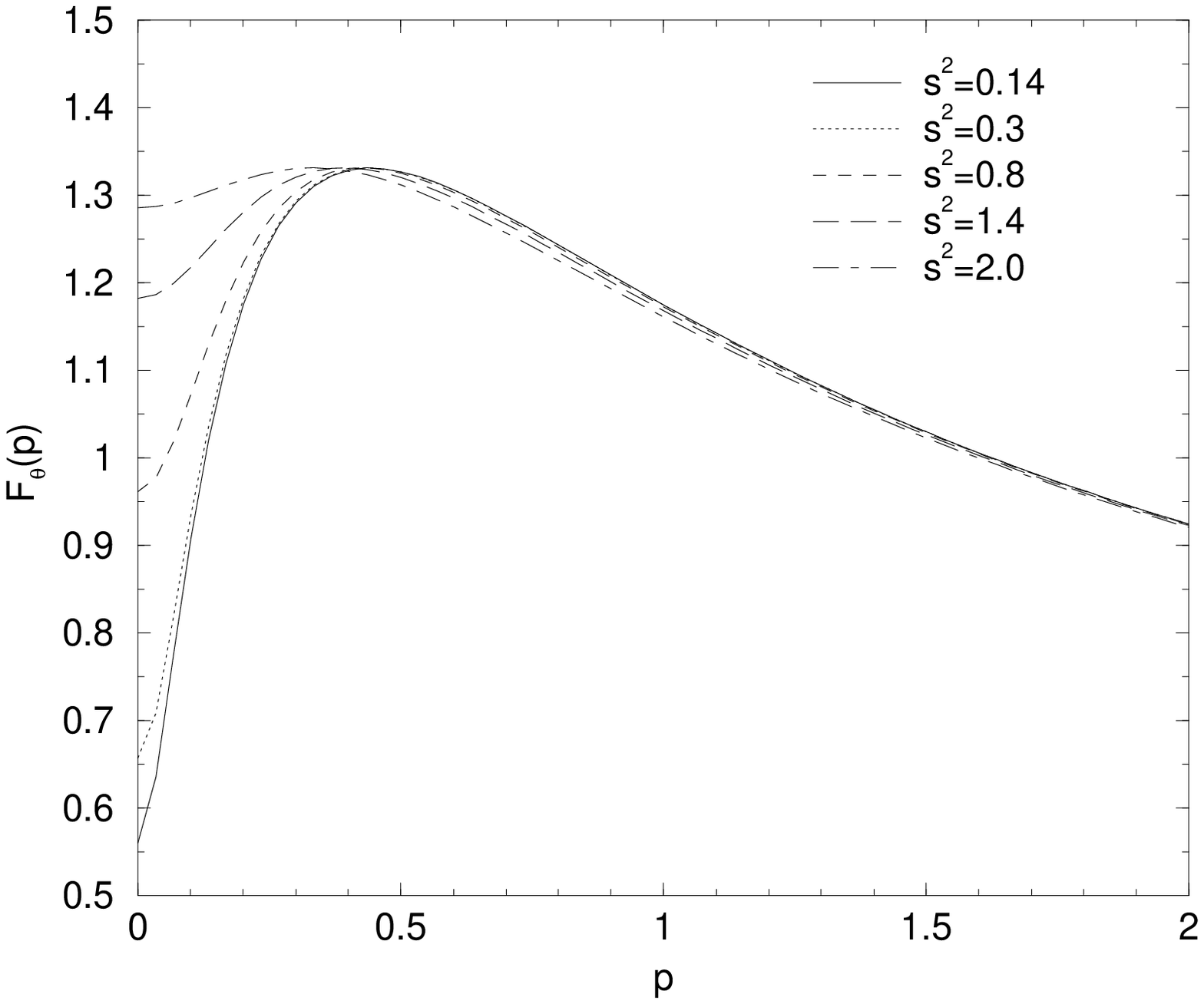}
\caption{
The RDA(24) enhancement factor as a function of $p$ for selected values 
of $s^2$.
}
\label{FtP1}
\end{figure}

After limited exploration, we used $i$$=$2, $j$$=$4. Again because the
motivating materials problem was brittle fracture in the presence of
water, our choice of training sets tended to focus on SiO.  We used
two molecules with Si--O bonds and two closed shell atoms,
$M=\{{\rm H_6Si_2O_7, H_4SiO_4, Be, Ne}\}$, with a set of six bond
lengths for each molecule.  That is, for the $\rm H_6Si_2O_7$ one of
the central Si--O bond lengths was changed, $R$(Si$_1$--O$_1$)$=$\{1.21,
1.41, 1.61, 1.91, 2.21, 2.81\} \AA. For $\rm H_4SiO_4$,
the deformation was in $T_d$ mode: all four Si--O bonds were changed
identically, $R$(Si--O$_i$)=\{1.237, 1.437, 1.637,
1.937, 2.237, 2.437\} \AA.
  KS-LDA densities and energies were the inputs (again see Appendix B for
computational details). Minimization of the target function defined by
Eq.~(\ref{IV1}) gave $b_2$$=$46.56873, $A_0$$=$0.51775, $A_1$$=$3.01873,
$\beta_1$$=$1.30030, $A_2$$=$$-$0.23118, and $\beta_2$$=$0.59016.
 A simple check shows
that the resulting enhancement factor $F_{\theta}^{\rm
RDA(24)}(\tilde\kappa_4)$ is positive for all positive values of
$\tilde\kappa_4$ (recall that $\tilde\kappa_4$ is positive by
definition).  Figures \ref{FtS2} and \ref{FtP1} show the
$F_{\theta}^{\rm RDA(24)}$ enhancement factor as a function of $s^2$ for
selected values of $p$ and, reciprocally, as a function of $p$ for
selected values of $s^2$.  This is a smooth, positive
function. $F_{\theta}^{\rm RDA(24)}(s^2,p\ge 0.4)$ becames practically
a straight line, essentially independent of $s^2$.

\begin{table}
\caption{\label{tab:table1}
KS kinetic energy $T_s$ values (in Hartrees) 
for selected molecules
and differences ($T_s^{\rm OF-DFT}-T_s^{\rm KS}$)
calculated using a GGA (Thakkar), MGGA, and RDA(24) explicit
semi-local approximate functionals. 
LDA-KS densities for LDA equilibrium geometries 
(calculated as described in Appendix B) were used as input.
} 
\begin{ruledtabular}
\begin{tabular}{lrrrr}
 & KS &  Thakkar & MGGA & RDA(24) \\
\hline
\hspace*{10pt}\\ [-8pt]
H$_2$            & 1.080  & $-$0.022 & 0.103 & 0.006 \\
LiH              & 7.784  &  0.021 & 0.296 & 0.063 \\
H$_2$O           & 75.502 & $-$0.285 & 0.318 & $-$0.128 \\
HF               & 99.390 & $-$0.353 & 0.329 & $-$0.148 \\
N$_2$            & 108.062 & $-$0.340 & 0.300 & $-$0.041\\
LiF              & 106.183 & $-$0.261 & 0.566 & 0.086 \\
CO               & 111.832 & $-$0.333 & 0.300 & $-$0.074 \\
BF               & 123.117 & $-$0.273 & 0.456 & 0.077 \\
NaF              & 260.097 & $-$0.348 & 1.295 & 0.648 \\
SiH$_4$          & 290.282 & 0.084  & 3.112 & 0.381 \\
SiO              & 362.441 & $-$0.262 & 2.825 & 0.293 \\
H$_4$SiO         & 364.672 & $-$0.163 & 3.338 & 0.293 \\
H$_4$SiO$_4$     & 587.801 & $-$0.860 & 4.133 & $-$0.034 \\
H$_6$Si$_2$O$_7$ & 1100.227 & $-$1.408& 7.968 & 0.086 \\
\hline
\hspace{20pt}\\[-6pt]
MAE\tablenotemark[1]              & ---      & 0.358    &  1.810 &  0.168\\
\end{tabular}
\end{ruledtabular}
\tablenotetext[1] {~~MAE$=$mean absolute error}
\end{table}

\begin{table}
\caption{\label{tab:table2}
Energy gradient (Hartree/{\AA}) calculated at point $R_m$ 
corresponding to the 
extremum of attractive force as calculated by the KS method.
Approximate OF-DFT energy gradients are obtained by 
replacing $T_s^{\rm KS}$ by $T_s^{\rm OF-DFT}$.
LDA-KS densities for LDA equilibrium geometries 
(calculated as described in Appendix B) were used as input.
} 
\begin{ruledtabular}
\begin{tabular}{lrrrrr}
 &  $R_m$, \AA & KS & Thakkar & MGGA & RDA(24) \\
\hline
\hspace*{10pt}\\ [-8pt]
H$_2$            & 1.2671 & 0.164 & 0.029 & 0.112 &  0.005 \\ 
LiH              & 2.455  & 0.046 & 0.016 & 0.037 & 0.016 \\ 
H$_2$O(1R)       & 1.3714 & 0.216 & $-$0.050& $-$0.073&  0.249\\
H$_2$O(2R)       & 1.3714 & 0.416 & $-$0.127& $-$0.163& 0.424 \\
HF               & 1.3334 & 0.232 & $-$0.071 & $-$0.003 & 0.180 \\
N$_2$            & 1.3986 & 0.576 & $-$0.349 & $-$0.819 & 0.244 \\
LiF              & 2.0405 & 0.079 & $-$0.019 & $-$0.032 & $-$0.007 \\
CO               & 1.4318 & 0.474 & $-$0.248 & $-$0.659 & 0.466 \\ 
BF               & 1.6687 & 0.207 & $-$0.037 & $-$0.118 & 0.204 \\ 
NaF              & 2.4284 & 0.067 & $-$0.007 & 1.169 & $-$0.008 \\
SiH$_4$          & 1.9974 & 0.447 & 0.102 & 0.189 & 0.101 \\
SiO              & 1.9261 & 0.278 & $-$0.098 & $-$0.281 & 0.175 \\
H$_4$SiO         & 2.057  & 0.162 & $-$0.027 & $-$0.086 & 0.151 \\
H$_4$SiO$_4$     & 2.037  & 0.712 & $-$0.278 & $-$0.714 & 0.745 \\
H$_6$Si$_2$O$_7$ & 2.010 & 0.194 & $-$0.022 & $-$0.173 & 0.165 \\
\end{tabular}
\end{ruledtabular}
\end{table}

Table \ref{tab:table1} displays kinetic energies for 14 molecules (four
of them with Si--O bonds) calculated at equilibrium geometries by
the conventional KS method and by approximate OF-DFT functionals using 
the KS density as input. The Thakkar empirical functional \cite{Thakkar92}
was chosen as an example of a GGA KE functional. The Perdew-Constantin
meta-GGA \cite{Perdew07} ``MGGA'' was chosen because it, like our
functional, is based on quantities that are at fourth order in the
density gradient expansion.
  The results are a bit  surprising, 
because the mean absolute error (MAE) for
the RDA functional is almost a factor of two smaller 
then the MAE for the Thakkar
KE, and almost ten times smaller then the MAE for the MGGA. Given the
parameterization to a small
training set containing only two molecules with Si--O bonds and two
closed shell atoms, we had not expected to obtain such 
good transferability to other systems.

Because our objective is a KE functional
capable of predicting correct interatomic forces, one of the 
important aspects is the behavior in the 
attractive regions of the potential surface.  Table \ref{tab:table2}
shows energy gradients for the molecules 
 in Table \ref{tab:table1} calculated at the stretched
bond length(s) for which the ``exact'' (i.e. reference) KS
atractive force has maximum magnitude.  One and two bonds were deformed in
the water molecule (respectively
denoted in the table as $\rm H_2O(1R)$ and $\rm H_2O(2R)$),
 while  $\rm SiH_4$ and $\rm H_4SiO_4$
were deformed in $T_d$ mode, and only one Si--O bond was stretched in
$\rm H_4SiO$ and $\rm H_6Si_2O_7$.

The forces were calculated by a three-point centered
finite-difference formula. As found in our previous work and 
summarized above, the
GGA functionals (with the Thakkar functional as the example GGA
functional here) are generally incapable of predicting the correct sign 
(attraction) for the
force; the only molecules in Table \ref{tab:table2} for which
GGA predicts attraction are H$_2$, LiH, and $\rm SiH_4$.
  We find the situation with the 
MGGA functional to be very similar; the predicted energy gradient has the
wrong sign in most cases.  In contrast, for all but two of the table entries
the RDA(24) functional
predicts the correct sign of the gradient.  In many cases ($\rm
H_2O(1R)$, $\rm H_2O(2R)$, HF, CO, BF, SiO, H$_4$SiO, $\rm H_4SiO_4$,
$\rm H_6Si_2O_7$) it yields values very close to the reference KS
results.

\begin{figure}
\epsfxsize=8.4cm
\epsffile{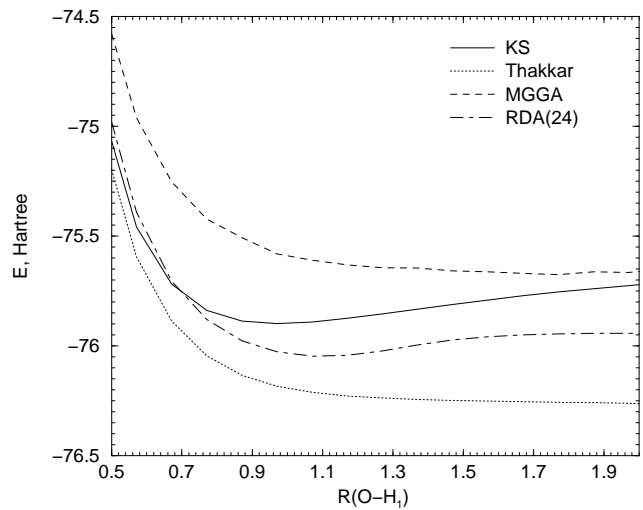}
\caption{
Total energy as a function of the O--H$_1$ distance for the H$_2$O
molecule (with O--H$_2$ kept at its equilibrium value)
obtained from a KS calculation with LDA XC
and from approximate GGA(Thakkar), MGGA(Perdew-Constantin)
 and RDA(24) functionals. 
The LDA KS densities (calculated as described in Appendix B)
 were used as input to the
orbital-free functionals.
}
\label{H2O-1R}
\end{figure}

Figure \ref{H2O-1R} shows the energy for the water molecule as a
function of the O--H$_1$ bond length.  Again, neither the GGA nor the MGGA
curve exibits a minimum.  This is a case for which the new RDA(24)
functional behaves relatively poorly.  It
 does  reproduce a minimum, but at too large a bond length,
while in the tail region its curve goes almost
flat. Thus the structure predicted by RDA(24) would be more expanded 
than the correct value and the attractive 
force in the tail region would be significantly underestimated.

\subsection{Atomic analysis of the RDA(24) functional}

For analysis of the new functional,  we calculated 
the Pauli potential near the nucleus
($r\rightarrow 0$) for the Be atom using a simple H-like density. 
A single-$\zeta$ Slater orbital
density with exponents $\zeta_{1s}=3.6848$ and
$\zeta_{2s}=0.9560$ taken from Ref.\ \onlinecite{ClementiRaimondi63} 
near $r=0$ has the following Taylor series expansion:
$n(r) \approx 415.0479 \times(1-7.3971~r)$.  The density
$n(r)=415.0479 \times \exp(-7.3971~r)$ has the same slope at $r=0$.
It can be used as an approximate density for the Be atom near $r=0$ to
calculate the Pauli potential for the RDA(24) Eq. (\ref{VI2}), for the GGA
\cite{TranWesolowski02} and for the PBE2 modified conjoint GGA
functionals.  Calculations were performed using our own {\sc Maple}
code. For RDA(24), we find
\bea v_{\theta}^{\rm RDA(24)}(r\rightarrow 0) %
&=&14010 - 5.9025\times 10^6~r \nonumber \\
&& +1.1271\times 10^9~r^2+{\rm O}(r^3) \,,
\label{VI3}
\eea
while for the GGA the result is 
\bea
v_{\theta}^{\rm GGA}(r\rightarrow 0) &=& \frac{-3.1961}{r}+ %
272.95 \nonumber \\
&& \!\!\!\!\!\!\! -1319.1~r+3257.5~r^2+{\rm O}(r^3) \, ,
\label{VI4}
\eea
and for the PBE2 modified conjoint GGA 
\bea
v_{\theta}^{\rm PBE2}(r\rightarrow 0) &= & %
\frac{0.742}{r}+ 265.31 \nonumber \\
&& \!\!\!\!\!\!\!\!\!\!\!-1319.6~r+3257.1~r^2+{\rm O}(r^3) \, .
\label{VI5}
\eea
As expected, the first term in the GGA potential is divergent 
and negative, while the PBE2 modified conjoint
GGA functional has a  divergent but positive first term. 
The numerical coefficients 
for the rest of the terms are very close for the GGA and PBE2
 modified conjoint 
GGA functionals.  This closeness is consistent with the analysis
in Ref.\ \onlinecite{Perdew92}.
\begin{figure}      
\epsfxsize=7.20cm
\vspace{0.5cm}
\centerline{\hspace*{+15pt}\epsffile{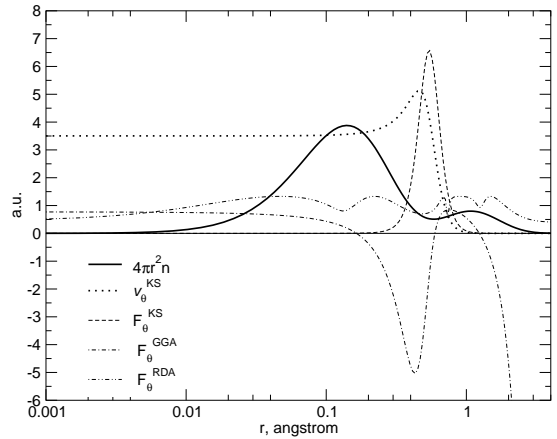}}
\vspace{1.1cm}
\epsfxsize=7.20cm
\centerline{\hspace*{+15pt}\epsffile{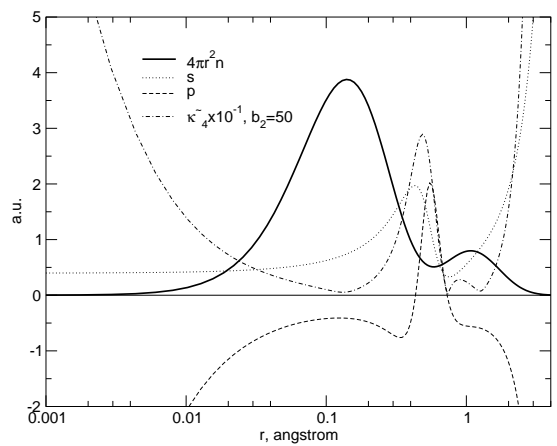}}
\vspace{0.3cm}
\caption{
(Upper panel): numerical KS-LDA density of the Be atom, 
KS Pauli potential and corresponding enhancement factor, and two approximate
enhancement factors.\\
(Lower panel): 
variables $s$, $p$, $\tilde\kappa_4$ ($b_2=50$) calculated for the Be atom using
KS-LDA density (also shown).
}
\label{Be}
\end{figure}

Figure \ref{Be} (upper panel) shows  the 
KS-LDA Pauli potential for the Be atom. 
Its value at the nucleus is approximately 3.5 Hartrees.
The RDA(24) potential has a finite (and positive) value at the nucleus, but
it is a strong overestimate; recall Eq.~(\ref{VI3}).
 Further comparison shows that 
the slope of the RDA(24) Pauli potential at the nucleus has a 
large negative value, whereas it should be 
very close to zero (see upper panel of Fig. \ref{Be}).
To complete the study, three enhancement factors, KS, GGA and RDA(24),
are shown in the same panel. Again, as was  seen in
Figure \ref{SiO-tv_theta-R2} for the SiO molecule, there is 
an important region where 
$F_{\theta}^{\rm GGA}$ is negative. The RDA(24) enhancement factor 
is positive 
eveywhere, but is still far from being an accurate approximation
to the KS form.  
$F_{\theta}^{\rm {RDA(24)}}$ 
has a sharp minimum near $ r \approx 0.75$ \AA~which 
corresponds to the change of sign of the reduced Laplacian of the 
density ($p$) 
(see the lower panel). The RDD $\tilde \kappa_4$ (shown in lower panel)
also has a sharp minimum near this point.

\section{Discussion and Conclusions} 

Success for the OF-DFT calculation of quantum forces in 
molecular dynamics requires a 
reliable explicit form for $T_{\rm s}$.  Though previously published
GGA-type (conjoint and nearly so) KE functionals yield reasonable KE
values, they fail to bind
simple molecules even with the correct KS density as input.
Therefore they produce completely  unusable interatomic forces.
This poor performance stems from violation
of the positivity requirement on the Pauli
potential.  Our first remedy was to constrain conjoint KE functionals
to yield positive-definite Pauli potentials.  Those functionals 
generate bound molecules and give semi-quantitative inter-atomic
forces. However, they are singular at the nuclear positions, hence severely
over-estimate the KS kinetic energy.  Examination of the near-nucleus
behavior of both original conjoint  and modified-conjoint GGA 
functionals shows that
the singularities cannot be eliminated within that simple
functional form.

Truncation of the gradient expansion, at higher orders in $s$ and $p$,
allows us to identify near-nucleus singular behavior and obtain
relationships among the coefficients of those truncations that will
eliminate such singularities.  The resulting
reduced density derivatives and related reduced-density-approximation
functionals are promising for the simultaneous description of kinetic
energies and interatomic forces.  

Two other aspects of the numerical results in Table \ref{tab:table1} 
relate to exact constraints, hence deserve
brief comment.  First, the von Weizs\"acker KE is the exact $T_s$ for
two-electron singlets.  We have not enforced that limit, yet the
error from RDA(24) in $H_2$ is only 6 mHartree.  Second, 
violation of $N$-representability by an approximate $T_s^{\rm appx}[n]$ 
is signaled by $T_s^{\rm appx}[n] - T_s[n] < 0$ for at least one $n$
\cite{Ayers07}.  
Five of the RDA(24) entries in Table \ref{tab:table1} have such  negative 
differences, while MGGA has none and the earlier GGA by Thakkar 
has many.  However, interpretation of those computed differences is 
a bit tricky, in that they do not correspond to the rigorous 
$N$-representability-violation test but to
$T_s^{\rm appx}[n] - T_s^{\rm LDA}[n]$.  With that limitation in mind,
the results in Table \ref{tab:table1} are at least suggestive of the
notion that the MGGA and RDA(24) functionals are $N$-representable
or, in some operational sense, close.  It is also important to
remember that the narrow goal is a functional that can be parameterized to
a small training set which is relevant to the desired materials simulations.
This limitation of scope is a practical means for limiting the risks
of non-$N$-representability.  In addition, 
we have many RDD forms open for exploration other than RDA(24).  

We close with a final word of caution.
The functional forms we have examined to this point, e.g. 
Eq.~(\ref{VI2}), may be too simple to provide robust and
transferable KE functionals for practical OF-DFT
applications. Moreover, the use of RDDs as basic variables in kinetic
energy enhancement factors guarantees the finiteness of the
corresponding Pauli potential only for those densities which satisfy
a generalization of Kato's cusp condition Eq.\ (\ref{V9bb}), and does 
not guarantee the satisfaction of the non-negativity property, Eqs.\
(\ref{TthetaPos})--(\ref{vthetaPos}).  The latter constraint must be
enforced separately.  Nevertheless the RDA scheme appears quite
promising and further development of it is underway.

\begin{acknowledgments}
We acknowledge informative conversations with Paul Ayers, Mel Levy,
Eduardo Lude\~na, John Perdew, Yan Alexander Wang, and 
Tomasz Wesolowski. This work 
was supported in part
by the U.S.\ National Science Foundation, grant DMR-0325553. FEH also
acknowledges support from NSF grant PHY-0601758.
\end{acknowledgments}

\begin{center}{\normalfont\small\bfseries
 APPENDIX A.~~FUNCTIONAL DERIVATIVES\\ OF $F_\theta$}
\end{center}

This appendix provides detail relative to derivation of the formulas
given in Eqs.~(\ref{vthetas2pq}), (\ref{V4a}), and (\ref{V4b}).
  All of them
follow from an evaluation of an expression of the generic form
presented as $v_\theta^{(2i)}({\bf r})$, Eq.~(\ref{V4}).  That
equation is a straightforward expression of the rules for the
evaluation of a functional derivative.  For clarity
in what follows, we restate here the definitions
\begin{equation}
s^2 = \frac{\nabla n \cdot \nabla n}{\xi^2 n^{8/3}}\, ,
~~ p=\frac{\nabla^2 n}{\xi^2 n^{5/3}}\, ,~~ t_0([n];{\bf r})
=c_0 n^{5/3} \, , \label{AppendixA2}
\end{equation}
where $\xi^2=4(3\pi^2)^{2/3}$ and
 $c_0=\frac{3}{10}(3\pi^2)^{2/3}$.
We also remind the reader that
\begin{eqnarray} &&
q=\frac{\nabla n \cdot \nabla\nabla n \cdot \nabla n}{\xi^4n^{13/3}},
\qquad q^\prime=\frac{\nabla n \cdot \nabla \nabla^2 n}{\xi^4 n^{10/3}},
\nonumber \\ [8pt] &&
q^{\prime\prime}=\frac{\nabla^4 n}{\xi^4n^{7/3}},
\qquad q^{\prime\prime\prime}=\frac{\nabla\nabla n : \nabla\nabla n}{\xi^4
n^{10/3}}; \label{AppendixA2a}
\end{eqnarray}
recall that $A:B \equiv \sum_{ij} A_{ij}B_{ji}$.  In addition, we introduce
the sixth-order reduced density derivatives
\begin{eqnarray} &&
h=\frac{|\nabla n \cdot \nabla\nabla n|^2}{\xi^6 n^6}\, , \qquad
h^\prime=\frac{|\nabla\nabla^2 n|^2}{\xi^6 n^4}\, , \nonumber \\ [8pt]
&&h^{\prime\prime}=\frac{(\nabla\nabla^2 n)\cdot
(\nabla\nabla n \cdot \nabla n)}{\xi^6 n^5}\, . \label{AppendixA2b}
\end{eqnarray}
Finally, we write $n$ instead
of the more explicit form $n({\bf r})$, and we
note that the derivatives of $s^2$ and $p$
with respect to $n$, $\nabla n$, and $\nabla^2 n$ are
\begin{displaymath}
\frac{\partial (s^2)}{\partial n}=-\frac{8s^2}{3n}\, , \quad
\frac{\partial (s^2)}{\partial (\nabla n)} = \frac{2\nabla n}{\xi^2 n^{8/3}}\, , \quad \frac{\partial (s^2)}{\partial (\nabla^2 n)} = 0\,;
\end{displaymath}

\vspace{-4pt}
\begin{equation}
\frac{\partial p}{\partial n}=-\frac{5p}{3n}\, , \quad
\frac{\partial p}{\partial (\nabla n)} = 0 \, , \quad
\frac{\partial p}{\partial (\nabla^2 n)} = \frac{1}{\xi^2 n^{5/3}}
\, . \label{AppendixA3}
\end{equation}
\hspace*{1pt}\\[-6pt]
Substituting these relations into Eq.~(\ref{V4}), and
restricting consideration to cases where $F_\theta$ depends only
on $s^2$ and $p$ (for which all the terms we need to use are
explicitly shown in that equation), we have immediately
\begin{eqnarray}
v_\theta({\bf r})&=& c_0 n^{2/3}\BIG{[}{14pt} \frac{5}{3} F_\theta
-\frac{8}{3} s^2 \left(\frac{\partial F_\theta}
{\partial (s^2)}\right)
-\frac{5}{3} p \left(\frac{\partial F_\theta} 
{\partial p}\right) \BIG{]}{14pt} \nonumber \\ [10pt]
&& \hspace{-20pt}-\,\frac{2c_0}{\xi^2}\,\nabla \cdot \BIG{[}{14pt}
\left(\frac{\partial F_\theta}{\partial (s^2)}\right) \frac{\nabla n}{n}
 \BIG{]}{14pt}
+ \frac{c_0}{\xi^2}\,\nabla^2\left(\frac{\partial F_\theta}
{\partial p}\right) \, . \label{AppendixA4}
\end{eqnarray}
At this point we remark that some terms that would otherwise be
expected in Eq.~(\ref{V4}) are absent because of the zeros in
Eq.~(\ref{AppendixA3}). 

\vspace{6pt}
To proceed further we need to expand the last two terms of
Eq.~(\ref{AppendixA4}). The first of these terms expands into
\begin{eqnarray} &&
-2c_0\, n^{2/3}\BIG{[}{14pt}\left(\frac{\partial F_\theta}
{\partial (s^2)}\right)\left\{
\frac{\nabla(n^{-1})\cdot \nabla n
 + n^{-1}\nabla^2 n}{\xi^2 n^{2/3}}
\right\} \nonumber \\ [8pt]
 &&  \hspace{0pt}+\,
 \left(\frac{\partial^2 F_\theta}{\partial (s^2)^2}
\right) \frac{\nabla s^2 \cdot \nabla n}{\xi^2 n^{5/3}} +
\left(\frac{\partial^2 F_\theta}{\partial p\, \partial (s^2)}\right)
\frac{\nabla p \cdot \nabla n}{\xi^2 n^{5/3}}\BIG{]}{14pt} \, .
 \hspace{20pt} \label{AppendixA5}
\end{eqnarray}
Observing now that $\nabla(n^{-1})=-n^{-2}\nabla n$ and that
\begin{eqnarray}
\nabla (s^2) &=& -\frac{8s^2}{3n}\nabla n + \frac{2\nabla n \cdot \nabla  
\nabla n}{\xi^2 n^{8/3}}\, , \nonumber \\ [8pt]
\nabla p &=& \frac{\nabla \nabla^2 n}{\xi^2 n^{5/3}} - \frac{5}{3}
\frac{(\nabla^2 n) \nabla n}{\xi^2 n^{8/3}} \, , \label{AppendixA8}
\end{eqnarray}
the expression in Eq.~(\ref{AppendixA5}) can be brought to the form
\begin{eqnarray}
&& +2c_0\,n^{2/3}\BIG{[}{14pt}
(s^2-p)\left(\frac{\partial F_\theta}{\partial (s^2)}\right)
\nonumber \\ [8pt] &&
+ \,\left(\frac{8}{3}s^4-2q\right)
\left(\frac{\partial^2 F_\theta}{\partial (s^2)^2}\right)
\nonumber \\ [8pt] &&
-\, \left(q^\prime-\frac{5}{3}s^2p\right)
\left(\frac{\partial^2 F_\theta}{\partial p\, \partial (s^2)}\right)
\BIG{]}{14pt} \, . \label{AppendixA9}
\end{eqnarray}

Continuing now to the final term of Eq.~(\ref{AppendixA4}), we expand the
Laplacian, obtaining initially
\begin{eqnarray}
\frac{c_0}{\xi^2}\,\nabla^2\left(\frac{\partial F_\theta}{\partial p}\right)
&=& c_0 n^{2/3}\BIG{[}{14pt}\left(\frac{\partial^2 F_\theta}{\partial(s^2)\,
\partial p}\right)\frac{\nabla^2 s^2}{\xi^2 n^{2/3}} \nonumber
\\ [8pt] && \hspace{-75pt}
+\left(\frac{\partial^2 F_\theta}{\partial p^2}\right)\frac{\nabla^2 p}
{\xi^2 n^{2/3}}
+\left(\frac{\partial^3 F_\theta}{\partial(s^2)^2\,\partial p}\right)
\frac{\nabla s^2 \cdot \nabla s^2}{\xi^2 n^{2/3}} \nonumber
\\ [8pt] && \hspace{-75pt}
+\,2\left(\frac{\partial^3 F_\theta}{\partial(s^2)\,\partial p^2}\right)
\frac{\nabla s^2 \cdot \nabla p}{\xi^2 n^{2/3}}
+\left(\frac{\partial^3 F_\theta}{\partial p^3}\right)\frac{\nabla p
\cdot \nabla p}{\xi^2 n^{2/3}} \BIG{]}{14pt}. \hspace{20pt}
\label{AppendixA10}
\end{eqnarray}

Using Eqs.~(\ref{AppendixA8}), we next find
\begin{widetext}
\begin{eqnarray}
\nabla^2 s^2 &=& -\frac{8}{3}\,\nabla \cdot
\left(\frac{s^2}{n}\nabla n\right) + \frac{2}{\xi^2}\,\nabla \cdot\left(
\frac{\nabla n \cdot \nabla \nabla n}{n^{8/3}}\right) \nonumber \\ [8pt]
&=& -\frac{8}{3}\left[\!-\frac{s^2}{n^2}\nabla n \cdot \nabla n
+\frac{\nabla(s^2)\cdot\nabla n}{n} +\,\frac{s^2}{n}\nabla^2 n\right]
+\frac{2}{\xi^2}\left[\!-\frac{8}{3}\frac{\nabla n \cdot \nabla\nabla n
\cdot \nabla n}{n^{11/3}}
+\frac{\nabla\nabla n : \nabla \nabla n + \nabla n \cdot \nabla \nabla^2 n}
{n^{8/3}}\right], \nonumber \\ [8pt]
\nabla^2 p &=& \frac{1}{\xi^2} \nabla \cdot \left(\frac{\nabla \nabla^2 n}
{n^{5/3}}\right) - \frac{5}{3\xi^2}\nabla \cdot \left(\frac
{(\nabla^2 n) \nabla n}{n^{8/3}}\right) \nonumber \\ [8pt]
&=& \frac{1}{\xi^2}\left[\!-\frac{5}{3}
\frac{\nabla \nabla^2 n \cdot \nabla n}{n^{8/3}}+\frac{
\nabla^4 n}{n^{5/3}}\right] - \frac{5}{3\xi^2}\left[
\!-\frac{8}{3}\frac{(\nabla^2 n)\nabla n \cdot \nabla n}{n^{11/3}}
+\frac{(\nabla^2 n)^2}{n^{8/3}} + \frac{\nabla\nabla^2 n \cdot \nabla n}
{n^{8/3}} \right], \nonumber \\ [8pt]
\nabla s^2 \cdot \nabla s^2 &=&
\frac{64s^4}{9n^2} \nabla n \cdot \nabla n - \frac{32s^2}{3}
\frac{\nabla n \cdot \nabla \nabla n \cdot \nabla n}{\xi^2 n^{11/3}}
+ 4\,\frac{|\nabla n \cdot \nabla\nabla n|^2}{\xi^4 n^{16/3}},
 \nonumber \\ [8pt]
\nabla s^2 \cdot \nabla p &=&
-\frac{8s^2}{3} \frac{\nabla n \cdot \nabla \nabla^2 n}{\xi^2 n^{8/3}}
+ 2\,\frac{(\nabla \nabla^2 n)\cdot (\nabla n \cdot \nabla
\nabla n)}{\xi^4 n^{13/3}} + \frac{40s^2}{9}\frac{(\nabla^2 n) \nabla n \cdot
\nabla n}{\xi^2 n^{11/3}} - \frac{10}{3}\frac{(\nabla^2 n)\nabla n \cdot
\nabla \nabla n \cdot \nabla n}{\xi^4 n^{16/3}},
\nonumber \\ [8pt]
\nabla p \cdot \nabla p &=&
\frac{|\nabla \nabla^2 n|^2}{\xi^4 n^{10/3}} - \frac{10}{3}
\frac{(\nabla^2 n)\nabla\nabla^2 n \cdot \nabla n}{\xi^4 n^{13/3}}
+\frac{25}{9}\frac{(\nabla^2 n)^2 \nabla n \cdot \nabla n}{\xi^4 n^{16/3}}.
\label{AppendixA11}
\end{eqnarray}

Then, combining material from Eqs.~(\ref{AppendixA4}),
 (\ref{AppendixA9}), (\ref{AppendixA10}), and
(\ref{AppendixA11}), and introducing the notations in
 Eqs.~(\ref{AppendixA2})--(\ref{AppendixA2b}), 
 we obtain the final result, applicable for any $F_\theta$
that depends only on $s$ and $p$:


\begin{eqnarray}
v_\theta &=& c_0\,n^{2/3}\BIG{[}{14pt} \frac{5}{3}F_\theta
-\left(\frac{2}{3}s^2+2p\right)
\left(\frac{\partial F_\theta}{\partial (s^2)}\right)
-\frac{5}{3}p \left(\frac{\partial F_\theta}{\partial p}\right)
 +\,\left(\frac{16}{3}s^4-4q\right)
\left(\frac{\partial^2 F_\theta}{\partial (s^2)^2}\right)
\nonumber \\ [8pt] &&
+ \, \left(\frac{88}{9}\,s^4+\frac{2}{3}\,s^2p - \frac{32}{3}\,q
+2\,q^{\prime\prime\prime}\right)\left(\frac{\partial^2 F_\theta}
{\partial (s^2)\partial p}\right) 
+ \, \left(\frac{40}{9}\,s^2p-\frac{5}{3}\,p^2-\frac{10}{3}\,q^\prime+
q^{\prime\prime}\right)\left(\frac{\partial^2 F_\theta}{\partial p^2}
\right)  \nonumber \\ [8pt] &&
+ \, \left( \frac{64}{9}\,s^6-\frac{32}{3}\,s^2q+4\,h\right)\left(
\frac{\partial^3 F_\theta}{\partial (s^2)^2 \partial p}\right)
+ \, \left(\frac{80}{9}\,s^4p-\frac{16}{3}\,s^2q^\prime-\frac{20}{3}\,p\,q
+4\,h^{\prime\prime}\right)\left(\frac{\partial^3 F_\theta}{\partial (s^2)
\partial p^2}\right) \nonumber \\ [8pt] &&
+\,\left(\frac{25}{9}\,s^2p^2-\frac{10}{3}\,p\,q^\prime+h^\prime\right)
\left(\frac{\partial^3 F_\theta}{\partial p^3}\right)\BIG{]}{14pt} \, .
\label{AppendixA12}
\end{eqnarray}
\end{widetext}

We may now specialize Eq.~(\ref{AppendixA12}) to the cases needed in
the present work.  Taking first $F_\theta^{\rm GGA}$, which has no $p$
dependence, all the terms of Eq.~(\ref{AppendixA12}) containing
derivatives with respect to $p$ vanish, leaving only the expression
previously given as Eq.~(\ref{vthetas2pq}).

Turning next to the specific forms of $F_\theta$ discussed in Section
IV-A, we note that
  $F_\theta^{\rm SGA}=1+a_2 s^2$ is not only independent of $p$, but
is also linear in $s^2$, so $\partial F_\theta/\partial (s^2)=a_2$ and
$\partial^2 F_\theta/\partial (s^2)^2=0$.  This causes $v_\theta^{\rm SGA}$
to have the form
\begin{equation}
v_\theta^{\rm SGA}=c_0\,n^{2/3}\BIG{[}{14pt}\frac{5}{3}(1+a_2s^2)
-a_2\left(\frac{2}{3}s^2+2p\right)\BIG{]}{14pt}\, ,
\end{equation}
which simplifies to the result given in Eq.~(\ref{V4a}).

Finally, we consider $F_\theta^{(4)}$ as given in Eq.~(\ref{V3}).
All the third derivatives of $F_\theta$  in Eq.~(\ref{AppendixA12}) vanish;
the first and second derivatives of $F_\theta$ have simple
forms.  We have
\begin{eqnarray}
v_\theta^{(4)} &=& c_0\,n^{2/3}\BIG{[}{14pt}\frac{5}{3}\left(a_4s^4+b_2p^2
+c_{21}s^2p\right) \nonumber \\ [8pt] &&
-\,\left(\frac{2}{3}s^2+2p\right)(2a_4s^2+c_{21}p)
\nonumber \\ [8pt] &&
-\,\frac{5}{3}p\,(2b_2p+c_{21}s^2)
+2a_4\left(\frac{16}{3}s^4-4q\right) \nonumber \\ [8pt] &&
+\,c_{21}\left(\frac{88}{9}\,s^4+\frac{2}{3}\,s^2p - \frac{32}{3}\,q
+2\,q^{\prime\prime\prime}\right) \nonumber \\ [8pt] &&
+\,2b_2\left(\frac{40}{9}\,s^2p-\frac{5}{3}\,p^2-\frac{10}{3}\,q^\prime+
q^{\prime\prime}\right) \BIG{]}{14pt}.
\label{AppendixA14}
\end{eqnarray}
Equation (\ref{AppendixA14}) simplifies to the result
 given as Eq.~(\ref{V4b}).

\begin{center}{\normalfont\small\bfseries
 APPENDIX B.~~COMPUTATIONAL METHODS}
\end{center}

We assess functionals by comparing results from them with
those of conventional orbital-dependent Kohn-Sham calculations
in the local density approximation (LDA), using standard methods
described in, for example,
Refs. \onlinecite{Slater2a,Slater2b,Slater2c,Gaspar,KohnSham65a,KohnSham65b,%
VWN80,CeperleyAlder80}.  
The reference molecular KS calculations were done with a triple-zeta
basis with polarization functions (TZVP) \cite{Ahlrichs92,Ahlrichs94,Basis}.
All integrals were calculated by a numerical integration scheme that,
following Becke \cite{Becke88}, uses weight functions localized near
each center to represent the multicenter integrals exactly as a sum of
(distorted) atomic integrals.  Radial integration of the resulting
single-center forms is accomplished by a Gauss-Legendre
procedure, while integration over the angular variables is done
with  high-order quadrature formulas developed by Lebedev and
coworkers \cite{Lebedev99a,Lebedev99b} with routines downloaded from
Ref.\ \onlinecite{CCL}.  These computations were performed using routines
developed by Salvador and Mayer \cite{SalvadorMayer04} and included in
their code {\sc fuzzy}.  The Vosko-Wilk-Nussair LDA \cite{VWN80} was used.

Given the KS density, for each OF functional under study we 
computed the total energy $E^{\rm OF\mbox{-}DFT}$ from
 Eq.~(\ref{A2}) and the
interatomic forces from Eq.~(\ref{A4}).
  The result is a non-self-consistent
calculation which tests whether a given OF functional can reproduce
$T_s[n_{KS}]$, or at least $\nabla_{\mathbf R} T_s[n_{KS}]$ if $n_{KS}$ is 
provided.  There is no sense in
trying to solve Eq.~(\ref{A3}) with an approximate OF functional
that cannot pass this test.



\begin{thebibliography}{99}

\bibitem{CarParrinello} For detailed information and publication references
 on Car-Parrinello Molecular Dynamics (CPMD), see URL
www.cpmd.org/cpmd.html.

\bibitem{GSA} 
D.E.~Taylor, V.V.~Karasiev, K.~Runge,
S.B.~Trickey, and F.E. Harris, Comp. Mater. Sci. {\bf 39}, 705
(2007).  

\bibitem{SM} F.H.~Streitz and J.W.~Mintmire, 
Phys. Rev. B {\bf 50}, 11996 (1994).

\bibitem{Goddard} M.J.\ Buehler, A.C.\ van Duin, and W.A.\ Goddard III, 
Phys. Rev. Lett. {\bf 96}, 095505 (2006).
 
\bibitem{Goddard2} A.K.~Rappe and W.A.~Goddard III,  J. Phys. Chem. {\bf 95},
3358 (1991).

\bibitem{QSDA} 
V.V.\ Karasiev, S.B.\ Trickey, and F.E.\ Harris,
Chem. Phys. {\bf 330}, 216 (2006).

\bibitem{Perspectives} 
V.V.\ Karasiev, S.B.\ Trickey, and F.E.\ Harris,
J. Comp.-Aided Mater. Des. {\bf 13}, 111 (2006).

\bibitem{Signpost} ``Recent advances in developing orbital-free
kinetic energy functionals'', V.V.~Karasiev, R.S.~Jones, S.B.~Trickey,
and F.E.~Harris, in {\it New Developments in Quantum Chemistry},
J.L.~Paz and A.J.~Hern\'andez eds. (Research Signposts), in press.

\bibitem{Hohenberg-Kohn}
P.\ Hohenberg and W.\ Kohn, Phys. Rev. B {\bf 136}, 864 (1964).

\bibitem{Yang-Parr-book}
R.G.\ Parr and W.\ Yang,  {\it Density Functional Theory of Atoms and
Molecules} (Oxford, New York, 1989).

\bibitem{DreizlerGrossBook}
R.M.\ Dreizler and E.K.U.\ Gross, {\it Density Functional Theory}
(Springer-Verlag, Berlin, 1990).

\bibitem{Thomas}
L.H.\ Thomas, Proc. Cambridge Phil. Soc. {\bf 23}, 542 (1927).
                                                                              
\bibitem{Fermi} 
E.\ Fermi, Atti Accad. Nazl. Lincei {\bf 6}, 602 (1927).

\bibitem{Weizsacker}
C.F.\ von Weizs\"acker, Z. Phys. {\bf 96}, 431 (1935).

\bibitem{LudenaKarasiev2002}
E.V.\ Lude\~na and V.V.\ Karasiev, 
in {\it Reviews of Modern Quantum Chemistry: 
a Celebration of the Contributions of Robert Parr}, edited by K.D.\ Sen 
(World Scientific, Singapore, 2002) pp. 612--665.

\bibitem{WangCarter2000}
Y. A.~Wang and E. A.~Carter, ``Orbital-free Kinetic-energy Density
Functional Theory'', Chap.~5 in {\it Theoretical Methods in
Condensed Phase Chemistry}, edited by S. D.~Schwartz (Kluwer, NY, 2000)
pp. 117--184.

\bibitem{Zhou06} 
B.-J.~Zhou and Y.A.~Wang, 
J. Chem. Phys. {\bf 124}, 081107 (2006).

\bibitem{Garcia0807}
D.~Garc\'{\i}a-Aldea  and J.E.~Alvarellos,
Phys. Rev. A {\bf 77}, 022502 (2008); 
J. Chem. Phys. {\bf 127}, 144109 (2007) and 
references in both.

\bibitem{Perdew07}
J.~P.~Perdew and L.A.~Constantin,
Phys. Rev. B {\bf 75}, 155109 (2007).

\bibitem{GarciaCervera08} C.J.\ Garcia-Cervera, 
Commun. Computat. Phys. {\bf 3}, 968 (2008).

\bibitem{Ghiringhelli08} L.M.\ Ghiringhelli and L.\ Delle Site,
Phys. Rev. B  {\bf 77},  073104  (2008)

\bibitem{Eek06} 
W.\ Eek and S.\ Nordholm, 
Theoret. Chem. Accounts  {\bf 115}, 266 (2006).

\bibitem{KS}
W.\ Kohn and L.J.\ Sham, Phys. Rev. {\bf 140}, A1133 (1965).

\bibitem{Teller} 
E.~Teller, 
Rev. Mod. Phys. {\bf 34}, 627 (1962). 

\bibitem{TalBader78}
Y.\ Tal and R.F.W.\ Bader, Int. J. Quantum Chem. {\bf S12}, 153 (1978).

\bibitem{BartolottiAcharya82}
L.J.\ Bartolotti and P.K.\ Acharya, J. Chem. Phys. {\bf 77}, 4576 (1982).

\bibitem{Harriman87}
J.E.\ Harriman, in {\it Density Matrices and Density Functionals}, 
R. Erdahl and V.H. Smith Jr. eds. (D. Reidel, Dordrecht, 1987), 359-373.

\bibitem{LevyOu-Yang88}
M.\ Levy, and H.\ Ou-Yang,
Phys. Rev. A {\bf 38}, 625 (1988).

\bibitem{Perdew92}
J.P.\ Perdew, Phys. Lett. A {\bf 165}, 79 (1992).

\bibitem{LeeLeeParr91}
H.\ Lee, C.\ Lee, and R.G.\ Parr, Phys. Rev. A {\bf 44}, 768 (1991).

\bibitem{Sham70} L.J.\ Sham, 
Phys. Rev. A {\bf 1}, 969 (1970)

\bibitem{LevyPerdewSahni84}
M.\ Levy, J.P.\ Perdew, and V.\ Sahni, Phys. Rev. A {\bf 30}, 2745 (1984).

\bibitem{Herring86}
C.\ Herring, Phys. Rev. A {\bf 34}, 2614 (1986).

\bibitem{Hodges}
C.H.\ Hodges, Can. J. Phys. {\bf 51}, 1428 (1973).

\bibitem{Gelfand-Fomin} I.M.~Gelfand and S.V.~Fomin, {\it Calculus
of Variations} (Prentice-Hall, Englewood Cliffs NJ, 1963), p.~42.

\bibitem{Korn}
G.A.\ Korn and T.M.\ Korn, {\it Mathematical Handbook for Scientists
and Engineers} (McGraw-Hill, NY, 1961).

\bibitem{Kato57}
T.~Kato, Commun. Pure Appl. Math. {\bf 10}, 151 (1957).

\bibitem{Bingel63} W.A.\ Bingel, 
Z. Naturforschung A {\bf 18}, 1249 (1963).

\bibitem{PackByersBrown66} 
R.T.\ Pack and W.B.\ Brown, 
J. Chem. Phys. {\bf 45}, 556 (1966)

\bibitem{March..VanDoren2000}
N.H.~March, I.A.~Howard, A.~Holas, P.~Senet, and V.E.~Van Doren,
Phys. Rev. A {\bf 63}, 012520 (2000). 

\bibitem{KryachkoLudena}
E.S.~ Kryachko and E.V.~ Lude\~na, {\it Energy Density Functional Theory
of Many-Electron Systems} (Kluwer, Dordrecht, 1990).

\bibitem{dens}
The nuclear-electron interaction for the point-nucleus model
$v_{\rm ne}^I({\bf r})=Z_I/|{\bf r}-{\bf R}_I|$ is the leading term in 
$v_{\rm KS} $ Eq. (\ref{A3})
in the limit ${\bf r}\rightarrow {\bf R}_I$.
All other terms in $v_{\rm KS}$ are finite at nucleus $I$ 
and hence are negligible in comparison to $v_{\rm ne}^I$.
Hence, the solution of Eq. (\ref{A3}) in the vicinity of ${\bf r}={\bf R}_I$ 
 is  a hydrogen-like density. 
Equation(\ref{denscusp2}) is appropriate also for molecular systems; see
Ref.\ \onlinecite{PackByersBrown66}.

\bibitem{Janosfalvi05} Zs.\ J\'anosfalvi, K.D.\ Sen, and \'A.\ Nagy,
Phys. Lett. A {\bf 344}, 1 (2005).

\bibitem{TranWesolowski02}
F.\ Tran and T.A.\ Wesolowski, Int. J. Quantum Chem. {\bf 89}, 441 (2002).

\bibitem{PerdewBurkeErnzerhof96}
J. P. Perdew, K. Burke, and M. Ernzerhof,
Phys. Rev. Lett. {\bf 77}, 3865 (1996).

\bibitem{AdamoBarone02}
C.\ Adamo and V.\ Barone, J. Chem. Phys. {\bf 116}, 5933 (2002).

\bibitem{LacksGordon93}
D.J.\ Lacks and R.G.\ Gordon, J. Chem. Phys. {\bf 100}, 4446 (1994).

\bibitem{DePristoKress87}
A.E.\ DePristo and J.D.\ Kress, Phys. Rev. A {\bf 35}, 438 (1987).

\bibitem{Thakkar92}
A.J.\ Thakkar, Phys. Rev. A {\bf 46}, 6920 (1992).

\bibitem{KH0001} R.A.~King and N.C.~Handy, 
Phys. Chem. Chem. Phys. {\bf 2}, 5049  (2000);
Mol. Phys. {\bf 99}, 1005 (2001).

\bibitem{BartolottiPrivCommun} Private communication, L.J.\ Bartolotti to
SBT

\bibitem{Murphy81}
D.R.~Murphy, Phys. Rev. A {\bf 24}, 1682 (1981).

\bibitem{Yang86}
W.~Yang, Phys. Rev. A {\bf 34}, 4575 (1986).

\bibitem{Yang..Lee86}
W.~Yang, R.G.~Parr, and C.~Lee, Phys. Rev. A {\bf 34}, 4586 (1986).

\bibitem{Ayers..Nagy01} P.W.~Ayers, R.G.~Parr, and A.~Nagy,
Int. J. Quantum Chem. {\bf 90}, 309 (2001).

\bibitem{ParrGhosh86} 
R.G.~Parr and S.K.~Ghosh, Proc. Natl. Acad. Sci. USA {\bf 83},
3577 (1986).

\bibitem{ClementiRaimondi63}
E.~Clementi and D.L.~Raimondi, J. Chem. Phys. {\bf 38},
2686 (1963).


\bibitem{Ayers07}
P.W.~Ayers and S.~Liu, 
Phys. Rev. A {\bf 75}, 022514 (2007).

\bibitem{Slater2a}
J.C.\ Slater, Phys. Rev. {\bf 81}, 385 (1951).

\bibitem{Slater2b} J.C.\ Slater,
Phys. Rev. {\bf 82}, 538 (1951).

\bibitem{Slater2c} J.C.\ Slater,
 J. Chem. Phys.  {\bf 43}, S228 (1965).

\bibitem{Gaspar}
R.\ G\'asp\'ar, Acta Phys. Hung. {\bf 3}, 263 (1954).

\bibitem{KohnSham65a}
W.\ Kohn and L.J.\ Sham, Phys. Rev. {\bf 140}, A1133 (1965).

\bibitem{KohnSham65b}
B.Y.\ Tong and L.J.\ Sham, Phys. Rev. {\bf 144}, 1 (1966).

\bibitem{VWN80}
S.H.\ Vosko, L.\ Wilk, and M.\ Nusair, Can. J. Phys. {\bf 58}, 1200 (1980).

\bibitem{CeperleyAlder80}
D.M.\ Ceperley and B.J.\ Alder, Phys. Rev. Lett. {\bf 45}, 566 (1980).

\bibitem{Ahlrichs92}
A.\ Sch\"afer, H.\ Horn, and R.\ Ahlrichs, J. Chem. Phys. {\bf 97}, 
2571 (1992)

\bibitem{Ahlrichs94}
A.\ Sch\"afer, C.\ Huber, and R.\ Ahlrichs, J. Chem. Phys. {\bf 100},
5829 (1994).

\bibitem{Basis}From the Extensible
Computational Chemistry Environment Basis Set
Database, Version 02/25/04,
Molecular Science Computing Facility, Environmental and Molecular
Sciences Laboratory, Pacific Northwest Laboratory,
P.O. Box 999, Richland, Washington 99352, USA, funded by the
U.S. Department of Energy (contract DE-AC06-76RLO).  See
http://www.emsl.pnl.gov/forms/basisform.html

\bibitem{Becke88}
A.D.\ Becke,
J. Chem. Phys. {\bf 88}, 2547 (1988).

\bibitem{Lebedev99a}
V.I.\ Lebedev and D.N.\ Laikov,
Dokl. Akad. Nauk {\bf 366}, 741 (1999).

\bibitem{Lebedev99b} 
V.I.\ Lebedev and D.N.\ Laikov,
Dokl. Math. {\bf 59}, 477 (1999).

\bibitem{CCL}
Computational Chemistry List (CCL) Archives:
http://www.ccl.net/

\bibitem{SalvadorMayer04}
P.\ Salvador and I.\ Mayer,
J. Chem. Phys. {\bf 120}, 5046 (2004).

\end{thebibliography}
\end{document}